\documentclass{emulateapj}

\newcommand{\Code}{\texttt{SEDONA}}

\newcommand{\kms}{\ensuremath{\mathrm{km~s}^{-1}}}
\newcommand{\Nifs}{\ensuremath{^{56}\mathrm{Ni}}}
\newcommand{\Cofs}{\ensuremath{^{56}\mathrm{Co}}}
\newcommand{\Fefs}{\ensuremath{^{56}\mathrm{Fe}}}
\newcommand{\msun}{\ensuremath{M_\odot}}
\newcommand{\texp}{\ensuremath{t_{\mathrm{exp}}}}
\newcommand{\delV}{\ensuremath{\Delta v_k}}

\newcommand{\Ilam}{\ensuremath{I_\lambda}}

\newcommand{\Slam}{\ensuremath{S_\lambda}}
\newcommand{\Jlam}{\ensuremath{J_\lambda}}

\newcommand{\Jbar}{\ensuremath{\bar{J}_\lambda}}

\newcommand{\BB}{\ensuremath{B_\lambda}}

\newcommand{\Adot}{\ensuremath{\dot{A}_{\mathrm{ph}}}}
\newcommand{\Edot}{\ensuremath{\dot{E}_{\mathrm{ph}}}}
\newcommand{\Edep}{\ensuremath{\dot{E}_{\mathrm{dep}}}}
\newcommand{\Edepi}{\ensuremath{\dot{E}^{i,j}_{\mathrm{dep}}}}
\newcommand{\PdV}{\ensuremath{{P_g\frac{\partial V}{\partial t}}}}

\newcommand{\thomsonCS}{\ensuremath{\sigma_\mathrm{T}}}
\newcommand{\alphaC}{\ensuremath{\alpha_\mathrm{c}}}
\newcommand{\alphaP}{\ensuremath{\alpha_\mathrm{p}}}
\newcommand{\StokesI}{\ensuremath{\mathbf{I}}}

\newcommand{\alphaExp}{\ensuremath{\alpha_{\mathrm{exp}}}}
\newcommand{\alphaAbsL}{\ensuremath{\alpha_{\mathrm{abs,line}}}}
\newcommand{\alphaAbsE}{\ensuremath{\alpha_{\mathrm{abs,exp}}}}
\newcommand{\alphaAbs}{\ensuremath{\alpha_{\mathrm{abs}}}}
\newcommand{\alphaff}{\ensuremath{\alpha_{\mathrm{ff}}}}
\newcommand{\alphabf}{\ensuremath{\alpha_{\mathrm{bf}}}}
\newcommand{\lamC}{\ensuremath{\lambda_{\mathrm{c}}}}

\shortauthors{Kasen, Thomas, \& Nugent}
\shorttitle{3-D Supernova Radiative Transfer}

\begin{document}
\bibliographystyle{apj}

\title{Time Dependent Monte Carlo Radiative Transfer Calculations For
3-Dimensional Supernova Spectra, Lightcurves, and Polarization}
\author{Daniel Kasen\altaffilmark{1,2}, R.C. Thomas\altaffilmark{3}, and
 P. Nugent\altaffilmark{3}}
\altaffiltext{1}{Allan C. Davis Fellow, Department of Physics and Astronomy, Johns Hopkins University,
Baltimore, MD 21218}
\altaffiltext{2}{Space Telescope Science Institute, Baltimore, MD 21218}
\altaffiltext{3}{Lawrence Berkeley National Laboratory, Berkeley, CA
94720}

\begin{abstract} 
We discuss Monte-Carlo techniques for addressing the 3-dimensional
time-dependent radiative transfer problem in rapidly expanding
supernova atmospheres.  The transfer code \Code\ has been developed to
calculate the lightcurves, spectra, and polarization of aspherical
supernova models.  From the onset of free-expansion in the supernova
ejecta, \Code\ solves the radiative transfer problem
self-consistently, including a detailed treatment of gamma-ray
transfer from radioactive decay and with a radiative equilibrium
solution of the temperature structure.  Line fluorescence processes
can also be treated directly. No free parameters need be adjusted in
the radiative transfer calculation, providing a direct link between
multi-dimensional hydrodynamical explosion models and observations.
We describe the computational techniques applied in \Code, and verify
the code by comparison to existing calculations.  We find that
convergence of the Monte Carlo method is rapid and stable even for
complicated multi-dimensional configurations.  We also investigate the
accuracy of a few commonly applied approximations in supernova
transfer, namely the stationarity approximation and the two-level atom
expansion opacity formalism.
\end{abstract}


\section{Introduction}

\subsection{Motivations}

Most of what we know about supernovae (SNe) has been learned from
observations of the lightcurve, spectrum, and polarization of the
supernova light during the months and years following the explosion.
Except for most nearby events, the explosion process itself is never
directly observed, and the progenitor star system only rarely.  What
we do see is emission from the hot, radioactive material ejected in
the explosion.  Theoretical radiative transfer modeling of the
emission is needed to discern the physical conditions in the SN
ejecta, offering insight into the physics of the explosion itself and
the progenitor star system which gave rise to it.

One dimensional (1D) explosion models of SNe have been used to
synthesize emergent spectra and light curves in reasonable agreement
with observed ones.  Because nearly all observed SNe are too distant
to be resolved in the early phases, deviations from spherical symmetry
can not be directly imaged.  Nevertheless, three lines of
observational evidence imply that the ejecta possess an interesting
multi-dimensional structure: (1) The detection of a non-zero intrinsic
polarization in SNe of all types, indicating a preferred direction in
the scattering medium
\citep[e.g.,][]{Cropper_87A,Kawabata_02ap,Wang_01el,Leonard_SNIa}; (2)
The appearance of unusual flux features in some SNe, naturally
explained by a clumpy ejecta structure, e.g., the ``Bochum event'' in
SN~1987A, \citep{Hanuschik_87A, Utrobin_87A}; (3) The complex
morphologies of SN remnants, which show clumpy, filamentary, or
jet-like structures \citep{Fesen_1996, Hwang_2000,Decourchelle_2001}.

Most current theoretical SNe explosion scenarios involve
multi-dimensional effects in an essential way.  In hydrodynamical
explosion models, ejecta asymmetries arise, for example, from random
instabilities in the explosion physics, e.g., Rayleigh Taylor
instabilities, convective mixing \citep{Chevalier-Klein, Burrows_1995,
Kifonidis_2000, Gamezo_3D, Roepke_fullstar}; from anisotropic energy
ejection mechanisms, e.g., jets, off-center ignition
\citep{khokhlov_jet, Macfadyen_1,Plewa_GCD}, or from asphericities in
the progenitor star or its surrounding medium, e.g., rapid rotation of
the progenitor, the presence of a binary companion star
\citep{Marietta,Yoon_2005}.

The multi-D explosion models make predictions as to the velocity,
composition and geometry of the material ejected in the SN explosion.
To confront such predictions with observations, the multi-D radiative
transfer problem in expanding SN atmospheres must be addressed.
Detailed radiative transfer codes synthesize model spectra, light
curves and polarization which can be compared directly to
observations.  Such transfer codes can also be usefully applied in an
empirical ``inverse'' approach, in which hand-tailored, parameterized
ejecta configurations are used to extract model-independent
information directly from the observations.

Here we describe a Monte Carlo approach to the multi-dimensional
time-dependent radiative transfer problem in expanding SN atmospheres,
embodied in the transfer code \Code.  Given an arbitrary 3-dimensional
ejecta structure (i.e., the density, composition and velocity
structure of freely expanding SN material) \Code\ self-consistently
calculates the emergent broadband light curves, spectral time-series
(in both optical and gamma-rays) and polarization spectra from various
viewing angles.  No free parameters need be adjusted in the transfer
calculations, providing a direct link between multi-dimensional
hydrodynamical explosion models and observations. In this paper, we
describe the radiative transfer techniques used in \Code, and give
some examples of its verification and application.

\subsection{Monte Carlo Radiative Transfer}

In the Monte-Carlo (MC) approach to radiative transfer, packets of
radiant energy (``photons'') are emitted from within the SN envelope
and tracked through random scatterings and absorptions until they
escape the atmosphere.  Each photon packet possesses a specific
wavelength and polarization state which are updated at each
interaction event.  Tallies of photon packets can be used to construct
estimators of the local radiation field properties and the emergent
spectrum at infinity.  The calculated quantities possess statistical
noise, which is reduced as the number of propagated packets is
increased.

The MC approach has several advantages over direct numerical solution
of the radiative transfer equation.  MC codes are intuitive,
relatively easy to develop, and generally less likely to fall victim
to subtle numerical errors \citep{Auer_MC}.  The method generalizes
readily to multi-dimensional time-dependent problems and the inclusion
of polarization.  As discussed in detail below
(\S\ref{rtm:Converge}), convergence of MC calculations is found to
be stable and rapid even for complicated configurations.  Finally,
although MC techniques can be computational expensive, they
parallelize well and can be profitably run on multi-processor
supercomputers.

Monte Carlo approaches have been applied to a wide range of
astrophysical radiative transfer problems, including multi-dimensional
polarization problems
\citep[e.g.,][]{Daniel_MC,Code_blobs,Wood_EScatterI}. The MC code
described in \cite{Hoeflich_93J} has been used to calculate the
continuum polarization and polarization spectrum of 2-D SN
\citep{Hoeflich_1991pol,Wang_96X,Howell_99by}.  In addition, the 1-D
MC code of \cite{Mazzali_MC} has been used in numerous studies of SN
flux spectra \citep[e.g.,][]{Mazzali_91T, Mazzali_98bw}.  The papers
of Leon Lucy have been particularly important in developing the MC
technique for astrophysical applications
\citep{Lucy_Radeq,Lucy_1999,Lucy1,Lucy2,Lucy3,Lucy_Timedep,Lucy_2005}.
The new techniques, many of which are applied here, make it feasible
for MC codes to match the physical accuracy of formal solutions of the
radiative transfer equation.

\section{Structure of the Radiative Transfer Code}

\subsection{Overview of Technique}
A short time after the eruption of a SN, hydrodynamic and
nucleosynthetic processes abate and the ejected material reaches a
phase of near free expansion.  Thereafter, the essential theoretical
challenge becomes to simulate the diffusion of photons through the hot
and optically thick ejecta -- i.e., the radiative transfer problem.
For epochs around and prior to peak brightness, the diffusion time of
photons exceeds the expansion time, such that the fully time-dependent
radiative transfer equation must be addressed.  In this case, MC
photon packets must be propagated through both space \emph{and} time.

The \Code\ code takes as input the density, composition, and shock
deposited energy specified at initial time $t_0$ at every point on a
spatial grid.  The spatial grid can be defined in a variety of
coordinate systems, including 1-D spherical, 2-D cylindrical, and 3-D
Cartesian systems.  In the free-expansion phase, the velocity of the
ejecta is homologous and everywhere proportional to radius: $r =
v\texp$ where $\texp$ is the time since explosion.  Given the
self-similar nature of the flow, velocity is used as the spatial
coordinates in the simulation, i.e., the spatial grid expands along
with the flow. 

A separate spatial grid exists for each time step in the model.  The
time discretization in the model most be chosen fine enough to resolve
the expansion of the SN ejecta.  Typically of order 100
logarithmically spaced time steps are used.  Following homology, the
density in a cell decreases with time as $(t/t_0)^{-3}$ while the
composition remains fixed.  The evolution of the local radiation field
and temperature in the cell are determined by the Monte Carlo
propagation of photon packets, as described below.

Propagation of the MC photon packets requires knowledge of the opacity
and emissivity of the SN ejecta at all points on the space-time
grid.  In general, these quantities depend upon the local radiation
field which heats and excites the gas.  Because the state of
the radiation field can only be known \emph{after} the MC simulation
has been run, an iterative approach is necessary to arrive at a
self-consistent model.  The overall iterative structure of the
transfer code is described as follows: \\
\begin{enumerate}
\item Using a 3-D gamma-ray transfer routine, we determine the rate of
energy deposition in each cell from the decay of radioactive \Nifs\
and \Cofs.  This, along with any initial shock deposited energy,
serves as the source geometry for the optical photon packets.
\item The opacities and emissivities at all wavelengths for each cell
and at each time step are computed.  Because the cell temperatures at
each time are not initially known, we start with a reasonable guess,
to be refined iteratively.
\item The propagation of optical photon packets through space and time
is followed, providing suitable tallies of the photon absorption rate
in each cell.
\item A new temperature is determined for each point on the space-time
grid by setting the rate of thermal emission equal to the calculated
rates of photon absorption plus any radioactive energy deposition.
\item The temperature structures calculated in step (4) will differ
from that used to compute the opacities in step (2).  Thus, to bring
about consistency, we recompute the opacities/emissivities and return
to step (3), iterating this procedure until the temperature and
opacities change negligibly from one iteration to the next.
\item Once the model atmosphere has converged, the synthetic
lightcurves and flux and polarization spectra are generated during
step (3) by collecting all photon packets escaping the atmosphere
along a certain line of sight.
\end{enumerate}

Before discussing each step in detail, we mention the important
physical approximations made in the present version of the code.
\begin{enumerate}
\item \emph{Homologous Expansion:} We assume the SN ejecta is in free
expansion, with a homologous velocity structure.  Thus the velocity
field of the ejecta is always spherically symmetric, even if the
ejecta density structure is not.  Free expansion is approached when
the ejecta have expanded sufficiently that the kinetic energy density
well exceeds the gravitational and internal energy densities. In
Type~Ia SN, this occurs less than a minute after the explosion
\citep{Roepke_Homo}; for Type~II SN, it can take as long as several
days \citep{Herant_Woosley}.  At later times, the energy input from
the decay of newly synthesized radioactive isotopes may produce
non-negligible deviations from homology \citep{Pinto-Eastman_II}.
Note that \Code\ does take into account adiabatic losses of the
radiation field, but in keeping with homology assumes that the ejecta
structure is negligibly affected by the energy exchange.  The current
calculations also neglect relativistic corrections going as $(v/c)^2$,
although these can be easily included if needed \citep{Lucy_2005}.

\item \emph{The Sobolev Approximation}: For atmospheres with large
velocity gradients such as SNe, the Sobolev approximation provides a
simple and elegant treatment of line transfer \citep{Sobolev_1947}.
Detailed derivations of the Sobolev formalism are given in
\citep{Mihalas_SA, Jeffery-Branch_1990, Castor_1970}.  The underlying
physical assumption is that the intrinsic profile of bound-bound
transitions is vanishingly narrow.  This is an excellent approximation
in SN atmospheres, in which the Doppler velocity width of lines ($v_d
\approx 5~\kms$) is typically much less than the velocity scale over
which the ejecta properties change ($v \approx 1000~\kms$).  Formal
inaccuracy may occur if the ejecta contain numerous small scale
structures or for very optically thick lines in which the Lorentz
wings become important.  In addition, \cite{Baron_RT} have emphasized
the problem that, given the enormous number of iron-peak lines at
ultraviolet wavelengths, several hundreds of lines may overlap within
a single Doppler width.  This overlapping clearly violates the
assumptions under which the Sobolev formalism is derived, although it
difficult to assess what sort of practical implications this has on
the transfer calculations.  The vast majority of the overlapping lines
are exceedingly weak, and the velocity spacing of optically thick
lines (which dominate the spectrum formation) is typically much larger
than a Doppler width \citep{Jeffery_Opacity}.  The errors thus
incurred on the emergent spectra are thus too small to notice, at
least in the few test calculations performed so far
\citep{Pinto-Eastman_Spectra}.  However, further head-to-head
comparisons (including non-equilibrium effects) are clearly needed.
For now, given the memory constraints of current computing facilities,
the Sobolev approximation appears unavoidable in multi-D
time-dependent calculations, for which the opacity of an enormous
number of lines must be stored on an extensive space-time grid.  In
this context, one anticipates any error incurred to cause
quantitative, not qualitative, variations in the emergent spectra and
lightcurves, and will likely not obscure the basic model dependencies
and orientation effects we are interested in studying.

\item \emph{Equilibrium Assumptions:} In the present models, we assume
the ionization/excitation state of the SN gas follows local
thermodynamic equilibrium (LTE) and can be calculated using the Saha
ionization and Boltzmann excitation equations.  We do not, however,
require the radiation field to be in equilibrium, and can include
scattering and fluorescence processes in the line source functions.
While the microscopic conditions for LTE (i.e., the dominance of
collisional rates) are not met in the rarefied atmospheres of SNe,
deviations of the atomic level populations from LTE should generally
cause quantitative, not qualitative differences in the emergent
spectra.  For Type~Ia SN, non-LTE effects are found to be small near
maximum light \citep{Baron_RT}, but become increasingly important
several months after the explosion.  A solution to the non-LTE rate
equations in the context of the Sobolev approximation is readily
incorporated into the MC approach \citep[see][]{Li-McCray_NLTE,
Zhang-Wang_NLTE, Lucy3} and future versions of \Code\ will include
such a solution for selected ionic species.
\end{enumerate}

\subsection{Calculation of Opacities and Emissivities}
\label{rtm:Opacity}

The important opacities in SN atmospheres are electron scattering,
bound-bound line transitions and, to a much lesser extent, bound-free
and free-free opacities.  With the temperature, density and
composition of the ejecta given, the LTE ionization and excitation of
the gas are determined by solving the Saha ionization and Boltzmann
excitation equations coupled to the equation of charge conservation.

Standard formulae for the extinction coefficients for electron
scattering and free-free opacities are found in
e.g.,~\cite{Rybicki_Lightman}.  We take bound-free opacities from
the Opacity Project \citep{Opacity_Project} when available, otherwise
the hydrogenic approximation is applied. All continuum opacities are
stored in discrete wavelength bins.

For the case of a single bound-bound transition, the extinction
coefficient is given by
\begin{equation}
\alpha_{bb} = K_{lu} \phi(\lambda),
\end{equation}
where $\phi$ is the line profile in the wavelength representation and
$K_{lu}$ is the (dimensionless) integrated line strength given by
\begin{equation}
K_{lu} = \biggl(\frac{\pi e^2}{m_e c}\biggr) f N_l  (\lambda_0^2/c)
\biggl(1 - \frac{N_u g_l}{N_l g_u} \biggr),
\end{equation}
where $f$ is the oscillator strength of the transition, $\lambda_0$
the line center rest wavelength, and $N_l$ and $N_u$ are the number
density of the lower and upper atomic levels respectively.  The last
term in parentheses is the correction for stimulated emission, where
$g_l$ and $g_u$ are the statistical weights of the lower and upper
atomic levels.

In a differentially expanding atmosphere, propagating photons are
continually Doppler shifting with respect to the local comoving frame.
The opacity of a bound-bound transition is thus only experienced when
a photon Doppler shifts into resonance with the line.  In the Sobolev
approximation, the spatial extent of the region of resonance is
assumed negligible, and the optical depth across the resonance region
is given by the Sobolev line optical depth
\begin{equation}
\tau = \frac{K_\lambda c \texp}{\lambda_0}.
\label{Eq:Sob_Tau}
\end{equation}
This simple equation only holds for atmospheres in homologous
expansion; for other velocity laws $\tau$ will depend upon the
direction the photon packet is traveling.  

The probability of a photon interacting with the line is $1 -
\exp(-\tau)$. In general, a photon will scatter multiple times in the
resonance region before escaping the line, where the escape
probability is given by
\begin{equation}
\beta = \frac{1 - e^{-\tau}}{\tau}.
\end{equation}
The conventional $\beta$ notation for the escape probability should not
be confused with the relativistic speed parameter $\beta = v/c$.

The result of the interaction of a photon with a line is the
redirection of the photon and its possible wavelength redistribution.
We consider three relevant atomic processes: pure scattering,
absorption/re-emission, and fluorescence.  The probability of the
photon being absorbed in the transition with lower level $l$ and upper
level $u$ is given by \citep{Pinto-Eastman_II}
\begin{equation}
p_{\mathrm{abs}} =  \frac{N_e \sum_k C_{uk}}
{N_e \sum_k C_{uk} + \sum_k \beta_{uk} A_{uk}} 
\label{Eq:P_Abs}
\end{equation}
here $N_e$ is the electron density, $C_{uk}$ the collision
coefficient, and $A_{uk}$ the Einstein spontaneous de-excitation
coefficient.  The sums runs over over all levels $k$ accessible from
the upper level $u$.  Collisonal coefficients can be calculated
approximately using Van Regemorter's formulae \citep{Van_Regemorter}.

For the conditions in SN atmospheres, the probability of true
absorption is found to be very small $p_\mathrm{abs} \approx
10^{-6}-10^{-4}$.  It is much more likely that the atom radiatively
de-excites and, more often than not, the de-excitation is a
fluorescence into an atomic level different than the original lower
level \citep{Pinto-Eastman_II}.

Atomic line data for the bound-bound transitions (including the
oscillator strengths and energy level data) have been taken from CD~23
and CD~1 of \cite{Kurucz_Lines}, containing over 500,000 and 40
million lines respectively.  Forbidden lines are not included in the
present calculations. In practice, it is often impossible to store the
optical depths for this many lines for all points on the space
time-grid.  Thus, for most calculation we select a subset (typically
from 0 to 500,000) of the most important lines to be given an
individual direct treatment with a more detailed approximation of
fluorescence.  All remaining lines are treated approximately by
combining them into a discrete opacity grid using the expansion
opacity formalism introduced by \cite{Karp} and later reformulated by
\cite{Pinto-Eastman_Spectra}
\begin{equation}
\alphaExp (\lamC) = 
\frac{1}{c \texp}
\sum_i \frac{\lambda_i}{\Delta\lamC}
(1 - e^{-\tau_i}),
\label{Eq:Exp_Opacity}
\end{equation}
where \lamC\ is the central wavelength of the bin and the sum runs
over all lines in the bin of width $\Delta \lambda_c$.  Preferably,
the wavelength bin sizes are $ \la 10~\AA$, to achieve reasonably well
resolved spectra.  The source function for these lines is treated in
the two-level atom (TLA) formulation
\begin{equation}
\Slam = (1 - \epsilon) \Jbar + \epsilon \BB(T).
\label{Eq:TLA_S}
\end{equation}
where $\epsilon$ represents the probability of absorption.  In
general, the value of $\epsilon$ is unique for each line and can be
calibrated by comparison to NLTE results \citep{Hoeflich_94D}.  In the
present case, we follow the approach of \cite{Nugent_hydro}, and
choose $\epsilon$ a common value for each line.  The validity of this
TLA approximation is investigated in \S\ref{sec:TLA}.

One can also define a purely absorptive component of the line expansion opacity
\begin{equation}
\alphaAbsE(\lambda) = \frac{1}{c \texp} \sum_i \frac{\lambda_i
}{\Delta\lamC} 
\biggl[\frac{\epsilon}{\beta + \epsilon(1-\beta)} \biggr]
(1 -
e^{-\tau_i}),
\end{equation}
as well as an purely absorptive component of the opacity from the
lines treated directly
\begin{equation}
\alphaAbsL(\lambda) = \frac{1}{c \texp} \sum_i \frac{\lambda_i
}{\Delta\lamC} p_{\mathrm{abs}} (1 - e^{-\tau_i}).
\end{equation}
The total absorptive opacity from all sources is
\begin{equation}
\alphaAbs = \alphaAbsL + \alphaAbsE + \alphabf + \alphaff,
\end{equation}
while the thermal emissivity from all sources is
\begin{equation}
j_\lambda(\lambda) = B(\lambda) \alphaAbs
\label{Eq:Emission}
\end{equation}

\subsection{Monte Carlo Transfer}

The optical photon packets used in the MC simulation are
monochromatic, \emph{equal energy} packets.  Each packet has initial
energy $E_p$ and represents a collection of $N_p = E_p \lambda /h c$
photons of wavelength $\lambda$.  The comoving frame energy of the
packet is kept fixed throughout any absorption/scattering interaction,
even if the packet wavelength is changed by the event.  The benefit of
this approach (as pointed out by \cite{Lucy_Radeq}) is that energy is
strictly conserved in each packet interaction.  This allows for rapid
convergence the correct temperature structure when the condition of
radiative equilibrium is imposed (\S\ref{rtm:Converge}).

\subsubsection{Emission of Photon Packets}
\label{Sec:Emission}
The luminosity of a SN is powered by the decay of radioactive isotopes
and/or thermal energy deposited by the SN shock-wave.  The primary
radioactive energy arises from the decay chain
\Nifs~$\rightarrow$~\Cofs~$\rightarrow$~\Fefs, with nearly all of the
decay energy emerging as $\sim 1$ MeV gamma-rays.  At early times,
these gamma-rays deposit energy in the ejecta primarily through
Compton scattering and photo-electric absorption.  Here we assume the
deposited energy is thermalized locally and instantaneously, although
non-thermal effects can readily be incorporated into the MC approach.

We determine the rate of radioactive energy deposition by following
the emission and propagation of gamma-rays using a MC routine
\cite[see][and references therein]{Milne_GR}.  Details of the routine
are given in the Appendix.  The gamma-ray transfer routine also
supplies the gamma-ray lightcurve and time-series of emergent
gamma-ray spectra as viewed from various inclinations.  These are
potentially powerful probes of the geometry and composition of the SN
ejecta \citep{Hungerford_GR, Hoeflich_3DGR}.

The calculated energy deposition rate in each cell along with the
inputted amount and distribution of internal shock deposited energy
existent at the initial time $t_0$ serve as the source of photon
packets.  All photon packets in the simulation (regardless of emission
time or location) are given the same initial energy $E_p$ in the
comoving frame, where
\begin{equation}
E_p = \frac{1}{N_p}
\biggl[ E_\mathrm{int,tot} + E_\mathrm{dep,tot} \biggr],
\end{equation}
where $E_\mathrm{int,tot}$ is the total amount of shock deposited and
$E_\mathrm{dep,tot}$ the total time-integrated radioactively deposited
energy. $N_p$ is the number of photon packets used in (each iteration
of) the simulation.  A fraction $E_\mathrm{int,tot}/E_p N_p$ of the
photon packets arise from the shock deposited energy and so are
emitted at the initial time $t_0$, and from a location sampled from
the spatial distribution of shock energy.  All remaining packets arise
from the radioactive energy deposition, and are emitted at a time and
location sampled from the instantaneous rate of energy deposition as
determined by the gamma-ray transfer procedure.  Emission is assumed
isotropic in the comoving frame, with the packet wavelength sampled
from the local thermal emissivity function (Eq.~\ref{Eq:Emission}).

Given the above prescription for packet emission, there is no need to
specify an inner boundary surface in the transfer simulation. Optical
photon packets are allowed to traverse the entire ejecta, even the
optically thick central regions.  This represents an improvement over
previous supernova MC codes \citep[e.g.,][]{Mazzali_MC} in which
photon packets are emitted from the surface of an extended spherical
inner core, with any packet backscattered onto the core assumed to be
``absorbed'' and removed from the calculation.

\subsubsection{Propagation of Photon Packets}
\label{rtm:Propogate}

Once emitted, photon packets are moved through the space-time grid in
small steps in velocity space. The propagation of packets resembles
that described in other MC studies \citep{Mazzali_MC, Lucy_Timedep}.
A velocity step of size $v$ corresponds to a physical distance $v
\texp$ and results in an elapse of time of $\delta t = \texp v/c$.  In
this way, packets are propagated through space and forward in time
until they reach the outer edge of the spatial grid, and which point
they are counted as being observed at the time of escape
(\S\ref{rtm:Emergent}).

Because of the differential expansion of the ejecta, the wavelength of
a propagating photon is continually Doppler shifting with respect to
the local comoving frame.  In a homologously expanding atmosphere,
this shift is always to the red and by an amount proportional to the
distance traveled: $\Delta \lambda = \lambda v/c$.  Photon packets
come into resonance with spectral lines one by one (given our
assumption of the Sobolev approximation) moving from blue to red.  The
velocity distance a packet propagates before coming into resonance
with a line treated individually (i.e., not in the expansion opacity
formalism) is
\begin{equation}
v_l = c (\lambda_0 - \lambda)/\lambda_0,
\end{equation}
where $\lambda$ is the comoving frame wavelength of the packet and
$\lambda_0$ is the rest wavelength of the line.  Meanwhile the
velocity distance a packet propagates before undergoing a continuum
interaction with the matter is determined randomly by
\begin{equation}
v_c = - \frac{1}{\alpha \texp} \log(z),
\end{equation}
where $\alpha$ is the total continuum opacity (i.e., the sum of the
bound-free, free-free, electron-scattering and line expansion
opacities) and $z$ is a random number uniformally sampled between 0
and 1, exclusive of the zero.  The next event for the packet is
determined by selecting the smallest value among $v_l, v_c,$ and
$v_s$, where $v_s$ is the shorter of the distance to the cell boundary
in the current direction of flight and the distance
$c(t_i-t_{current})$ where $t_i$ is the next time boundary.

If a continuum interaction occurs, the possible fate of the packet is
an absorption, electron scattering, or expansion-opacity line
scattering.  The nature of the event is determined by randomly
sampling the local scattering and absorption fractions.   An absorbed
packet is assumed thermalized and immediately re-emitted with a new
wavelength sampled from the local thermal emissivity
(Eq.~\ref{Eq:Emission}).  For scattered packets, the comoving
wavelength remains unchanged.  Electron scattering differs from line
expansion opacity scattering in its effect on the packet's
polarization state (see \S\ref{rtm:Pol_Calcs}).

When a packet comes into resonance with a line being treated directly
(i.e., not binned into the expansion opacity) an interaction occurs if
\begin{equation}
z < 1 - \exp(-\tau).
\end{equation}
If an interaction does occur, the packet will either be absorbed (with
probability given by Equation~\ref{Eq:P_Abs}) or will radiatively
de-excite.  In the case of radiative de-excitation, the probability
that the packet de-excites to atomic level $j$ is
\begin{equation}
p_{\mathrm{uj}} = \frac{ \lambda_j^{-1} \beta_{uj} A_{uj}} {\sum_k
\lambda_k^{-1} \beta_{uk} A_{uk}},
\end{equation}
where the Einstein $A$ coefficients of each line have been weighted by
the escape probabilities $\beta$ (in order not to count those
emissions that do not escape the line resonance region and almost
immediately re-excite the atom) and by the inverse wavelength of the
lines in order to get the energy distribution of the fluorescence
correct.

For every interaction event, a new propagation direction of the packet
is chosen randomly from an isotropic distribution (except in the case
of electron scattering, see \S\ref{rtm:Pol_Calcs}).  The new outgoing
rest frame energy is
\begin{equation}
E_{\mathrm{out}} = E_{\mathrm{in}} \frac{1 - \mu_{\mathrm{in}} v/c}{1 - \mu_{\mathrm{out}} v/c},
\label{Eq:Adi_Loss}
\end{equation}
where $\mu_{\mathrm{in}}$ and $\mu_{\mathrm{out}}$ are the cosines of
the angles between the photon propagation direction and the radial
direction for the incoming and outgoing packet respectively.
Equation~\ref{Eq:Adi_Loss} accounts for adiabatic energy losses of the
radiation field on a scatter-by-scatter basis.  In keeping with
homology, we assume that the energy exchange has a negligible effect
on the ejecta structure.

At the earliest times, the ejecta opacities are so large that
diffusion processes are insignificant. Then it is not necessary to
follow packets through numerous scatters.  Packets emitted at time $t$
prior to a chosen start time of $t_1 \approx 2$~days are held in place
until $t_1$, suffering an adiabatic energy loss over this time by a
factor $t/t_1$.

\subsubsection{Polarization Calculations}
\label{rtm:Pol_Calcs}
The calculation of polarization is readily incorporated into the Monte
Carlo approach.  Each photon packet is now assigned a Stokes vector
which describes the electric field intensity along two perpendicular
axes which are themselves perpendicular to the propagation direction
\begin{eqnarray}
\bf{I} = \left( \begin{array}{c} I\\Q\\U\end{array} \right)
= 
\left( \begin{array}{c}
I_{0^\circ} + I_{90^\circ} \\
                I_{0^\circ}  - I_{90^\circ  }\\
                I_{45^\circ} - I_{-45^\circ}
\end{array} \right)
\label{Eq:stokes_eq}
\end{eqnarray}
where $I_{90^\circ}$, for instance, designates the intensity measured
90$^\circ$ counterclockwise from a specified reference direction when
facing a source.  A fourth Stokes parameter $V$ measuring the circular
polarization is neglected here. For scattering atmospheres without
magnetic fields, the radiative transfer equation for circular
polarization separates from the linear polarization equations,
allowing us to ignore $V$ in our calculations \citep{Chandra_1960}.

In choosing a polarization reference axis for a packet moving in
direction $\vec{D}$, we use the following convention: consider the
plane defined by $\vec{D}$ and the z-axis; the reference axis is
chosen to lie in this plane and perpendicular to $\vec{D}$.  To
transform the Stokes vector to another reference axis rotated by an
angle $\psi$ clockwise, one applies the rotation matrix
\citep{Chandra_1960}
\begin{eqnarray}
R(\psi) = 
\left(
\begin{array}{ccc}
       1        &         0       &      0     \\
       0        &      \cos2\psi   &   \sin2\psi \\
       0        &     -\sin2\psi   &   \cos2\psi \\
\end{array}
\right).
\label{Eq:Pol_Rot_Matrix}
\end{eqnarray}

The thermal emission within the SN envelope is the result of random
collisional processes and hence unpolarized.  Photon packets are
thus initially assigned an unpolarized Stokes vector normalized to unity:
$\StokesI = (1,0,0)$.  The effect of an electron scattering on the
Stokes vector is described by application of the Rayleigh phase matrix
\begin{eqnarray}
P(\Theta) = \frac{3}{4}
\left( \begin{array}{ccc}
\cos^2 \Theta + 1     &   \cos^2 \Theta - 1    &   0 \\
\cos^2 \Theta - 1     &   \cos^2 \Theta + 1    &   0 \\
     0                &         0              & 2 \cos \Theta \\
\end{array}\right)
\label{Eq:Rayleigh_Matrix}
\end{eqnarray}
where $\Theta$ is the angle between the incoming and the scattered
photon.  Note that, given the generality of the MC approach, one is
not restricted to using only this particular phase matrix, but any
general polarizing effect can easily be considered.

The Rayleigh phase matrix of Equation~\ref{Eq:Rayleigh_Matrix} only
applies when the Stokes vectors are referred to the plane of
scattering.  More generally, the effect on a packet Stokes vector is
given by
\begin{equation}
\StokesI_{\mathrm{out}} = R(\pi - i_2) P(\Theta) R(-i_1) \StokesI_{\mathrm{in}}.
\label{Eq:ES_Pol}
\end{equation}
The rotation matrix $R(i_1)$ rotates the incoming packet Stokes vector
onto the scattering plane, while $R(\pi - i_2)$ rotates the outgoing
packet Stokes vector back into our conventional reference axis.  The
rotation angles $i_1$ and $i_2$ can be determined from the geometry 
\citep[see][]{Chandra_1960}. 

When polarization is taken into account, the intensity of electron
scattered radiation is not isotropic.  After each scatter we choose
new direction angles by sampling the anisotropic redistribution
implied by Equation~\ref{Eq:ES_Pol} using a standard rejection method
\citep{Code_blobs}.  The Stokes vector is always renormalized to unity.

Light scattered in a bound-bound line transitions may be polarized in
a way similar to electron scattering.  For a resonance line, the
polarizing effect can be expressed by the hybrid phase matrix derived
by \cite{Hamilton_1947}.  In addition, one must take into account
multiple scattering of photons within the line resonance region, which
can be treated analytically using the Sobolev-P formalism of
\cite{Jeffery_SobolevP}, who employs a hybrid phase matrix for all
lines as a crude approximation to the actual polarizing behavior.  For
optically thick lines, this multiple scattering tends to randomize the
directionality and hence depolarize the average emission from a
resonance region.  In addition, collisions between electrons tends to
randomly redistribute the atomic state of excited atoms over all the
nearly degenerate magnetic sublevels, thereby destroying the
polarization information \citep{Hoeflich_93J}.  For these reasons, the
polarizing effect of lines is not expected to be important in SN
atmospheres, and we typically assume that line scattered light is
unpolarized.

\subsection{Calculation of the Temperature Structure}
\label{rtm:Temp_Calc}

\begin{deluxetable*}{rrrrrrrrrrr} 
\tablewidth{0pt}
\tablecaption{Model parameters used in Example Calculations}
\tablehead{
Model  
 & $t_1$\tablenotemark{a}
 & $\log dt$\tablenotemark{b}
 & $n_t$\tablenotemark{c}
 & $n_r$\tablenotemark{d}
 & $n_x$\tablenotemark{e} 
 & $v_{\rm max}$\tablenotemark{f}
 & $n_\lambda$\tablenotemark{g} 
 & $\mathrm{max}_\lambda$\tablenotemark{h} 
 & $N_p$\tablenotemark{i} 
 & $\epsilon$\tablenotemark{j} }
\startdata
Lucy      &  2 &  0.015   &  100  &  200 &   -  &  10000 & 2000  &  20000  &  1e8 &  - \\
SYNOW     &  18  &  -      &    1   &  100 &  - &  20000 & 10000 &  30000 &  1e7 &  0 \\
w7-1D LC  &  2  &  0.0175 &  100  &  100  &   - &  30000 & 10000 &  30000  &  1e8 &  1  \\
w7-1D spec  &  18  &  - &  1  &  100       &   -  &  30000 & 10000 &  30000  &  1e6 &  1  \\
w7-2D spec  &  18  &  -      &    1  & 200 &  100 &   30000 & 10000 & 30000  &  1e7 &  1  \\ 
\enddata
\tablenotetext{a}{$t_1$: start time of time grid (in days)}
\tablenotetext{b}{$\log~dt$: logarithmic spacing of time grid (in days)}
\tablenotetext{c}{$n_t$: number of time points used in grid}
\tablenotetext{d}{$n_r$: number of radial zones}
\tablenotetext{e}{$n_x$: number of azimuthal zones, in 2-D cylindrical calculations}
\tablenotetext{f}{$v_{\rm max}$: Outer boundary of the spacial grid (\kms)}
\tablenotetext{g}{$n_\lambda$: number of wavelength groups}
\tablenotetext{h}{$\mathrm{max}_\lambda$: maximum wavelength (in Angstroms)}
\tablenotetext{i}{$N_p$: Number of optical photon packets used in each iteration}
\tablenotetext{j}{$\epsilon$: TLA absorption probability}
\label{Tab:grid}
\end{deluxetable*}

Calculation of temperature at each point in the SN ejecta is
necessary to determine the opacities and emissivities of the gas.
The evolution of the gas energy density $e$ in a volume $V$ is
governed by first law of thermodynamics,
\begin{equation}
\frac{\partial (e V)}{\partial t} = V(\Adot - \Edot + \Edep) - \PdV,
\label{Eq:energy_balance}
\end{equation}
where \Adot\ is the rate of optical/UV photon absorption, \Edot\ is
the thermal photon emission rate, and \Edep\ the rate of heating
from the decay of radioactive isotopes (all quantities in
ergs~s$^{-1}$~cm$^{-3}$).  The \PdV\ term is the rate of adiabatic
cooling, where $P_g$ is the gas pressure.

For the epochs of interest here (a few days to a few months after
explosion) the energy density is heavily radiation dominated, and the
time-scale for the thermalization of radioactive gamma-ray energy and
the absorption and emission of optical photons ($t_r = 1/c
\alpha_{\mathrm{abs}}$, where $\alpha_{\mathrm{abs}}$ is the absorption
opacity) is short compared to the other time-scales in the problem, in
particular the expansion time \texp\ of the ejecta, and the diffusion
time of photons.  Therefore a quasi steady-state is reached such that
the terms with time derivatives in Equation~\ref{Eq:energy_balance}
can be dropped \citep{Pinto-Eastman_I}.  For example, the ratio of the
\PdV\ term to the $V \Adot$ term is seen to be
\begin{equation}
\frac{\PdV}{V \Adot} \sim 3 \biggl[\frac{P_g}{a T^4}\biggr]
\biggl[ \frac{t_r}{\texp} \biggr],
\end{equation}
where we have used $V \propto \texp^3$ (since we assume homologous
expansion) and $\Adot\approx a T^4/t_r$.  Both terms in brackets are
$\ll 1$ for the reasons given above.  For example, for conditions
appropriate for the inner layers of ejecta of a SNe~Ia near maximum
light (number density $N = 10^9$; $T = 15000$~K, $\alpha = 10^{-14}$)
one finds $P_g/a T^4 \approx 5\times 10^{-6}$ and $t_r/\texp \approx
0.002$.  Thus the \PdV\ term can be neglected without incurring much
error.  An essentially similar argument can be made for the
$\frac{\partial (e V)}{\partial t}$ term.  We conclude that the gas
temperature reaches an equilibrium on a very short time-scale, and
Equation~\ref{Eq:energy_balance} can to good accuracy be taken as
\begin{equation} 
\Adot  + \Edep = \Edot.
\label{Eq:qs_energy_balance}
\end{equation}
Equation~\ref{Eq:qs_energy_balance} says that at each point, and at
each moment in time, heating of the gas by gamma-ray and photon
absorption is exactly balanced by cooling by thermal emission. Thus, a
solution of the differential Equation~\ref{Eq:energy_balance} to
determine the thermal evolution of the gas is not necessary (although
such a solution could readily be incorporated if a higher level of
accuracy is desired).  We emphasize that
Equation~\ref{Eq:qs_energy_balance} does not amount to a neglect of
all time-dependent effects. For the radiation field, adiabatic losses
and ejecta expansion are both relevant over a diffusion time, and
both are included naturally in the time-dependent propagation of MC
photon packets (\S\ref{rtm:Propogate})

One determines the heating terms in
Equation~\ref{Eq:qs_energy_balance} from the MC simulation. The rate
of radioactive energy deposition \Edep\ is known from the gamma-ray
transfer procedure (\S\ref{Sec:Emission}).  The rate of photon
absorption \Adot\ can be estimated from the MC transfer by counting
the energy of photon packets passing through a cell.  When a packet
with comoving frame energy $E_j$ takes a step of size $v_j$ in a cell,
the contribution to the absorbed energy is
\begin{equation}
dE  =  E_j \alphaAbs v_j \texp.
\end{equation}
Summing over all packet steps that occur in the cell during the MC transfer
routine gives
\begin{equation}
\Adot = 
\frac{E_p}{V_c~\Delta t} 
\sum_j \alphaAbs v_j \texp \frac{E_j}{E_p},
\label{Eq:Abs}
\end{equation}
where $V_c$ is the volume of the cell and $\Delta t$ is the elapsed
time covered by this grid time-slice.  As discussed by
\cite{Lucy_Radeq}, using the analytic estimator Equation~\ref{Eq:Abs}
is superior to simply counting the number of absorption events that
occur in a cell, as all packets moving though a cell contribute to the
calculation of \Adot, regardless of whether absorption occurs or not.
Thus Equation~\ref{Eq:Abs} remains a good estimator even with the
absorption probability in a cell is very low.

With the heating terms thus specified,
Equation~\ref{Eq:qs_energy_balance} can be solved for the new net
thermal emission, which satisfies
\begin{equation}
\Edot(T_{\mathrm{new}}) = 4\pi \int_0^\infty \alphaAbs(\lambda,T_{\mathrm{old}})
B(\lambda,T_{\mathrm{new}}) d\lambda,
\label{Eq:T_Equation}
\end{equation}
where we have made use of the LTE approximation for the emissivity as
a function of wavelength \citep[e.g.][p. 26]{Mihalas_SA} and $B$ is
the Planck function.  A new temperature $T_{\mathrm{new}}$ is
determined by solving Equation~\ref{Eq:qs_energy_balance} with
Equation~\ref{Eq:T_Equation} at each point on the space-time grid.  An
explicit solution for $T_{\rm new}$ can be derived
\begin{equation} T_{\rm new}=
{h\over k}\left[ (\Adot + \Edep)/(4\pi)
\over (2h/c^{2}) \int_{0}^{\infty} \alpha_{\rm abs}(x,T_{\rm old})
\left({x^{3}\over e^{x}-1}\right)\,dx \right]^{1/4} \,\, ,
\end{equation} where $x=hc/(kT\lambda)$.
The resulting time-dependent temperature structure will generally
differ from the initial temperatures $T_{\mathrm{old}}$ used to
compute the opacities and emissivities used in the MC transfer
procedure.  The model must thus be iterated until convergence is
reached, i.e., at each point on the grid, $T_{old} \approx
T_{\mathrm{new}}$ to acceptable accuracy.  Convergence is achieved
simultaneously on the entire space-time grid, rather than converging
each individual epoch as time advances.  The subject of convergence is
discussed in more detail in \S\ref{rtm:Converge}.

\subsection{The Emergent Spectrum}
\label{rtm:Emergent}
The final emergent spectra and lightcurves of the SN model are easily
obtained by collecting all escaping photon packets.  Escaping packets
are binned in time of arrival, observed wavelength, and escape
direction (i.e., viewing angle).  A large number of bins in each
dimension must be used to achieve the requisite resolution, and enough
packets collected in each bin to provide adequate photon statistics.
Broad-band lightcurves are constructed by convolving the spectrum at
each time with the appropriate filter transmission functions.

One can improve upon the packet collection method using a number of
variance reduction techniques, for example formal integral techniques
\citep{Lucy_1999, Mazzali_MC}.  These offer tremendous gains in
computational efficiency, especially in multi-dimensional problems, as
in this case all photon packets are used in the construction of the
spectrum, not just those escaping along a particular line of sight.

\section{Implementation and Verification}
\label{rtm:Test_Calcs}

The \Code\ radiative transfer code has been written in C++, and
parallelized using a hybrid of MPI and OpenMP.  For problems in which
the memory requirements are not exceedingly large, parallelization is
trivial and scaling perfect.  Each processor merely propagates its own
set of photon packets, with only minimal communication required to
combine the results at the end.

In \Code, the spatial model grid is defined in velocity coordinates,
such that the grid expands naturally with the ejecta.  The spatial
grid can be defined in a variety of coordinate systems, including 1-D
spherical, 2-D cylindrical, and 3-D Cartesian systems.  The photon
packets themselves are always tracked in real 3-D Cartesian
coordinates, and the current location on the grid is determined after
each packet step by mapping the packet coordinates onto the cell
geometry.  Defining additional coordinate systems, even irregular
ones, is easily accomplished, requiring only a new mapping function.

The time discretization in the model most be chosen fine enough to
resolve the expansion of the SN ejecta.  Typically, $\sim 100$ time
steps are used, logarithmically spaced and beginning at start time
$t_1$ of usually 2 days and ending at $\sim 100$~days. For all
calculations discussed below, the grid dimensions and other model
parameters used are given in Table~\ref{Tab:grid}.

\Code\ has been run on as many as 1024 processors at once on the AIX
IBM SP supercomputer Seaborg at NERSC, and has been tested on parallel
Mac and Linux platforms as well.  Numerous verification tests have
been performed, a few of which we discuss below.

\subsection{LightCurve Calculations}
\label{Sec:LC_test}

\begin{figure}[t]
\begin{center}
\includegraphics[width=8.5cm,clip=true]{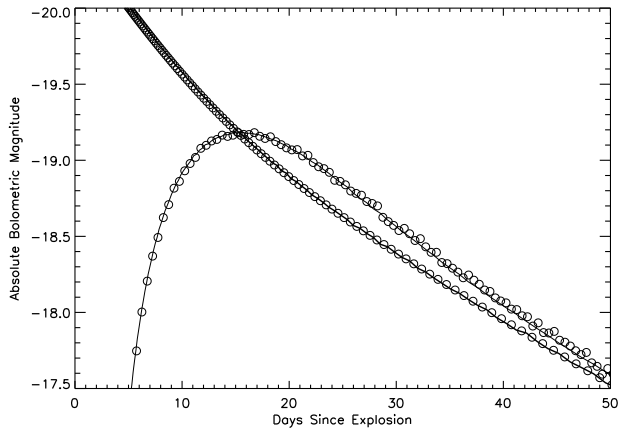}
\caption{\Code\ lightcurve calculation (open circles) for the test
SNe~Ia model discussed in \S\ref{Sec:LC_test} compared to the
solution of the comoving frame equations presented in
\cite{Lucy_Timedep} (solid lines).  Good agreement is found for both
the rate of gamma-ray energy deposition from radioactive decay
(exponentially declining curve) and in the observer-frame bolometric
lightcurve.
\label{Fig:Lucy_LC}}
\end{center}
\end{figure}

\begin{figure}
\begin{center}
\includegraphics[width=8.5cm,clip=true]{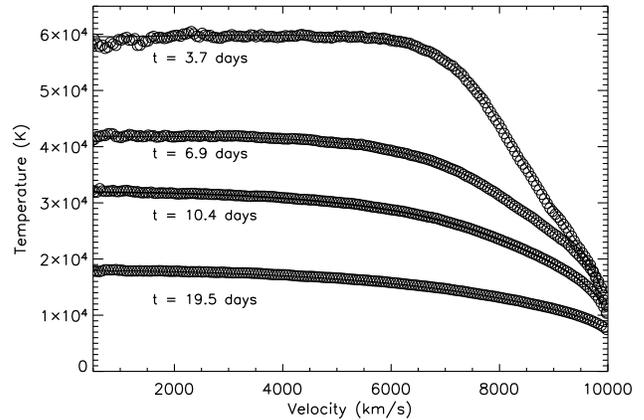}
\caption{\Code\ calculation of the temperature structure (open
circles) at a few select times for the test SN~Ia model, compared to
the numerical results presented in \cite{Lucy_Timedep} (solid lines).
\label{Fig:Lucy_Temp}}
\end{center}
\end{figure}

\begin{figure}[t]
\begin{center}
\includegraphics[width=8.5cm,clip=true]{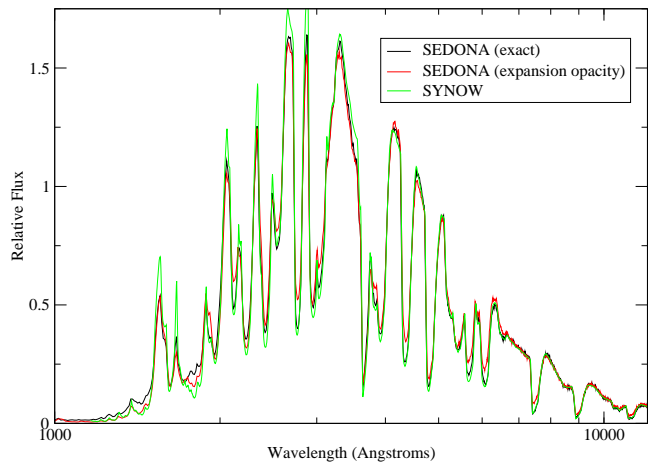}
\caption{\Code\ synthetic spectra of a pure silicon atmosphere
compared to the output of the SYNOW code.  
\label{Fig:synow_fig}}
\end{center}
\end{figure}

\cite{Lucy_Timedep} discusses a simple spherically-symmetric SN~Ia
model used to test lightcurve calculations.  The model consists of 1.4
\msun\ of constant density ejecta extending to 10000 \kms.  The \Nifs\
abundance is unity for the inner 0.5 \msun\ of ejecta, then drops
linearly to zero at 0.75 \msun, yielding a total \Nifs\ mass of 0.625
\msun.  For the test calculations, a grey absorption coefficient of
$\alpha/\rho= 0.1$~g~cm$^{-2}$ is adopted.

We have calculated a spherical \Code\ lightcurve for the model using
the parameters given in the first row of Table~\ref{Tab:grid}.
Although this problem is one of monochromatic grey radiative transfer,
to fully test the MC transfer procedure and temperature solver we use
2000 opacity wavelength bins each of which is set to the same (purely
absorbing) grey opacity.  The results of the \Code\ calculation
(Figure~\ref{Fig:Lucy_LC}) show excellent agreement with the numerical
solution of the comoving moment equations presented in
\cite{Lucy_Timedep}.  Detailed agreement obtains for both the rate of
gamma-ray energy deposition from radioactive decay and the
observer-frame bolometric light curve.  This verifies both our
gamma-ray deposition procedure and our primary MC transfer routine.
The discrepancy in the light curves is comparable to the error
attributed to the Eddington approximation in the comoving frame
calculations by \cite{Lucy_Timedep}.

We have further tested the \Code\ calculations of the ejecta
temperature structure using the same model.
Figure~\ref{Fig:Lucy_Temp} shows that, at all epochs, our radial
temperature structure is in excellent agreement with the numerical
results of \cite{Lucy_Timedep}.  This confirms that the \Code\ MC
estimators obtain the correct mean intensity of the radiation field in
the time-dependent transfer calculation.

\subsection{Spectrum and Polarization Tests}

\begin{figure*}
\begin{center}
\includegraphics[width=18.0cm,clip=true]{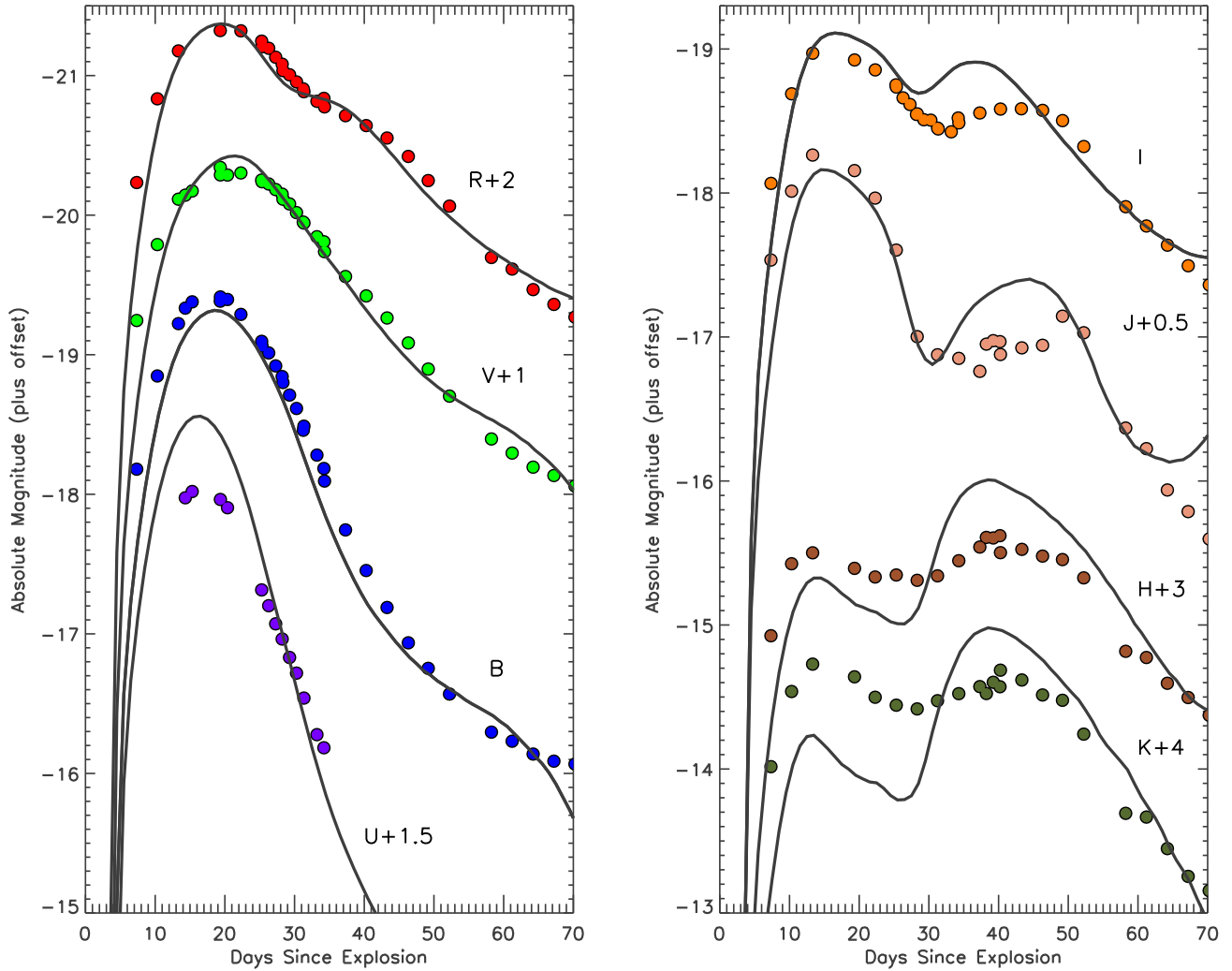}
\caption{\Code\ calculation of the UBVRIJHK lightcurves of the w7
Type~Ia supernova explosion model.  Overplotted are the
\cite{Kris_01el} observations of SN~2001el, corrected for extinction.
\label{Fig:W7_LC}}
\end{center}
\end{figure*}

\begin{figure}
\begin{center}
\includegraphics[width=8.5cm,clip=true]{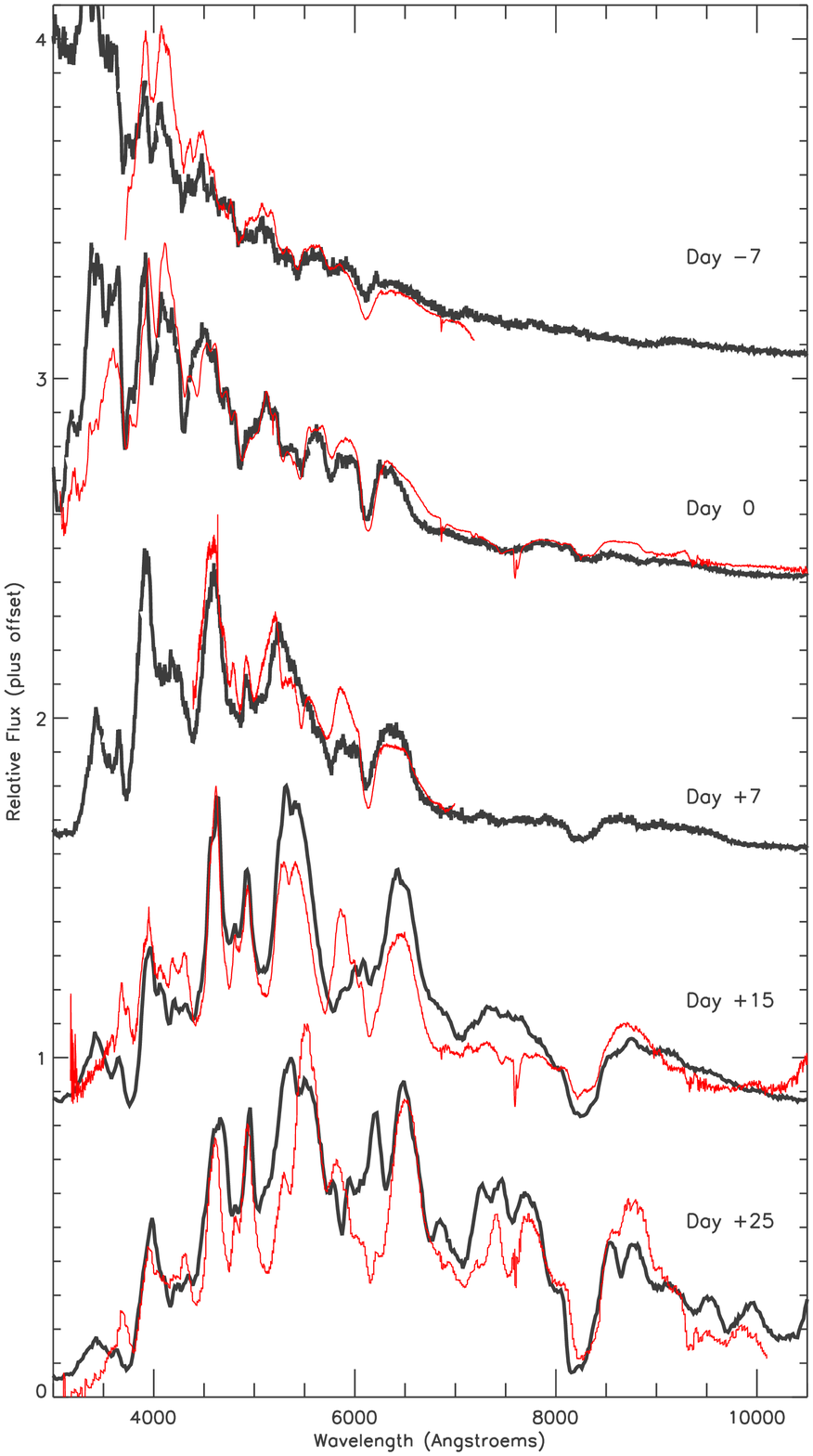}
\caption{\Code\ calculation of the spectra at several epochs of the w7
model compared to observations of SN~1994D.  Dates given are days
relative to B-band maximum.
\label{Fig:W7_Spectra}}
\end{center}
\end{figure}

Basic spectrum formation in \Code\ has been tested by comparing to
SYNOW, a 1-D spectrum synthesis code used in many analyses of SN
spectra \citep[][and references therein]{Branch_SNIb}. SYNOW solves
the transfer equation assuming the Sobolev approximation,
time-independence, a perfectly sharp blackbody emitting photosphere,
no continuum opacity, and a pure scattering line source function.
Test synthetic spectra were computed with \Code\ under the same
assumptions and the model parameters given in row 2 of
Table~\ref{Tab:grid}.  The particular test model discussed here
consisted of pure silicon ejecta extending to 20000~\kms\ with a
constant density ($\rho = 10^{-13}$~g/cm$^3$) and temperature ($T =
10000$~K).  The photosphere was located at 10000~\kms.

Figure~\ref{Fig:synow_fig} compares the SYNOW spectrum to two \Code\
spectrum calculations -- one in which the line opacity is treated
directly, and another in which all the lines have been binned into the
quasi-continuous expansion opacity (Eq.~\ref{Eq:Exp_Opacity}).  In
each case, the lines are assumed to be pure scattering --
i.e. fluorescence is ignored and $\epsilon = 0$.  

For the direct line treatment, the agreement of the spectrum with that
of SYNOW is quite good, implying a proper reproduction of the line
source function for blended lines.  Differences of the order a few
percent are noticed, and can be attributed to the fact that different
linelist data is used in the separate codes.  The expansion opacity
spectrum, on the other hand, does show noticeable discrepancies in the
depth of individual line features.  This is clearly a failure of the
formalism to properly treat wavelength bins that contain only one very
optically thick line ($\tau_l \gg 1)$.  According to
Eq.~\ref{Eq:Exp_Opacity}, the optical depth accrued in redshifting
across such a bin is unity, and the probability of interacting with
the line 1 - $\exp(-1) \approx 0.63$.  This underestimates the true
interaction probability $1 - \exp(-\tau_l) \approx 1$.  A modification
to the expansion opacity formalism to better handle this limit may be
warranted.

\Code\ calculations of the emergent continuum polarization have been
tested as well for a variety of cases.  In the optically thin limit,
we have verified the pure electron scattering polarization levels
against the semi-analytical formulae of \cite{Brown-Mclean}.  In the
optically thick case, we have reproduced the plane-parallel results of
\cite{Chandra_1960} and those of the axisymmetric configurations
calculated in \cite{Hillier_1994}.

\subsection{Example Application Calculations}
\label{rtm:Example}

As an example of the application of \Code\ to a realistic research
problem, we calculate the spectra and lightcurves of the parameterized
1-D SN~Ia explosion model w7 \citep{Nomoto_w7,Thielemann_w7}.  Several
previous radiative transfer studies using w7 have shown the model to
be reasonable -- but not perfect -- accordance with observations
\citep{Branch_w7,Harkness_w7,Jeffery_1992, Nugent_hydro, Lentz_94D}.
Note that some earlier 2-D, time-independent \Code\ spectral
calculations based upon w7 have already appeared
\citep{Kasen_hole,Kasen_GCD}.

Figure~\ref{Fig:W7_LC} shows the synthetic UBVRIJHK lightcurves from
the \Code\ calculation of w7 using the parameters in row 3 of
Table~\ref{Tab:grid}.  For these calculations we have used the Kurucz
CD1 linelist containing over 40 million lines.  All lines are treated
in the expansion opacity two-level atom formalism with $\epsilon = 1$
(i.e., no lines are given a direct fluorescence treatment).
Overplotted in Figure~\ref{Fig:W7_LC} are observations of the normal
Type~Ia SN~2001el \citep{Kris_01el} assuming a distance modulus of
31.45 and corrected for an extinction of $A_v = 0.5, R_v = 2.88$. The
broadband model lightcurves resemble those of the observations in most
regards, such as the rise times, decline rates, and peak magnitudes.
A clear secondary maximum is seen in the model near-infrared IJHK
bands, although it is generally stronger than the observations.  These
lightcurves can be compared to w7 transfer calculations using the
STELLA code \citep{Blinnikov}, which show a similar behavior in the
optical bands.

Figure~\ref{Fig:W7_Spectra} compares the computed w7 spectra to
observations of the Type~Ia SN~1994D \citep{Patat_94D,Meikle_94D} at
several epochs.  Given the low estimated dust extinction to this
object, no redding correction has been applied. While exact agreement
is not to be expected, given the known limitations of the w7 model, on
the whole the model well reproduces the general spectral features and
colors of the observations.  The overall sensible behavior suggests
that time-dependent \Code\ calculations can be used to model the
spectroscopic evolution of SNe over a wide time span.

At yet later times ($\texp \ga 70$~days), SN~Ia spectra become
increasingly nebular, with NLTE effects and cooling by forbidden
emission lines playing dominate roles.  Perhaps most signficantly, at
these times one expects non-thermal ionization due to the products of
radioactive decay to become important once the temperature drops low
enough that LTE predicts neutrality \citep{Swartz_1991}.  This physics
is not included in the present version of \Code, and the emergent
spectra and broadband lightcurves should not be expected to be
accurate at the latest epochs.  However, note that \cite{Branch_94D}
has shown, surprisingly, that resonance line scattering by permitted
transitions actually does a good job of characterizing the spectrum
out past \texp = 200 days.

\subsection{Convergence Properties}
\label{rtm:Converge}

\begin{figure}
\begin{center}
\includegraphics[width=8.5cm,clip=true]{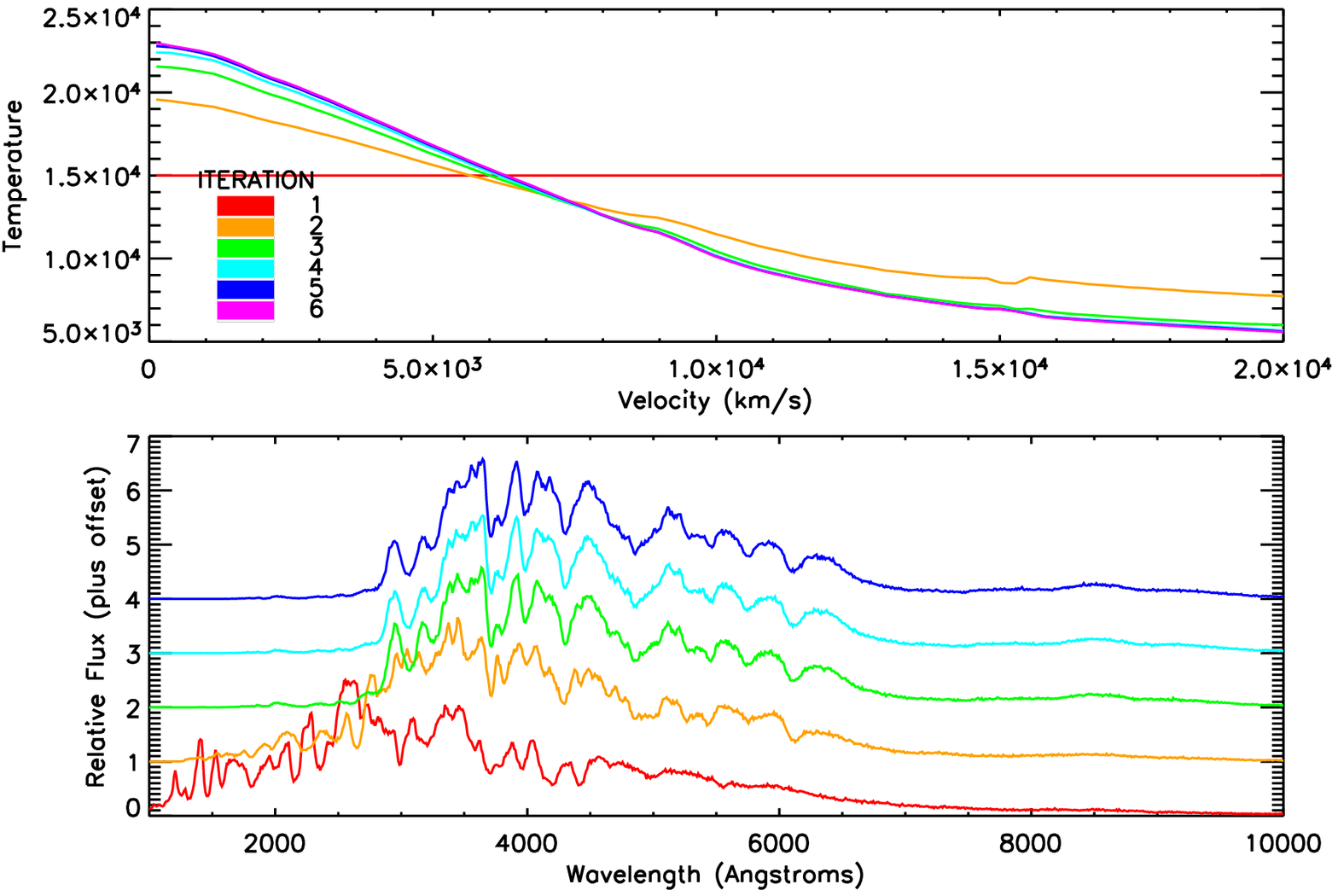}
\caption{Convergence of a day 18 w7 spectrum model, showing the
variation of the radial temperature structure (top panel) and emergent
spectrum (bottom panel) with iteration.  Beginning with an isothermal
temperature structure, the model converges to better than 1\% in only
5 iterations.
\label{Fig:w7_converge}}
\end{center}
\end{figure}

\begin{figure}
\begin{center}
\includegraphics[width=8.5cm,clip=true]{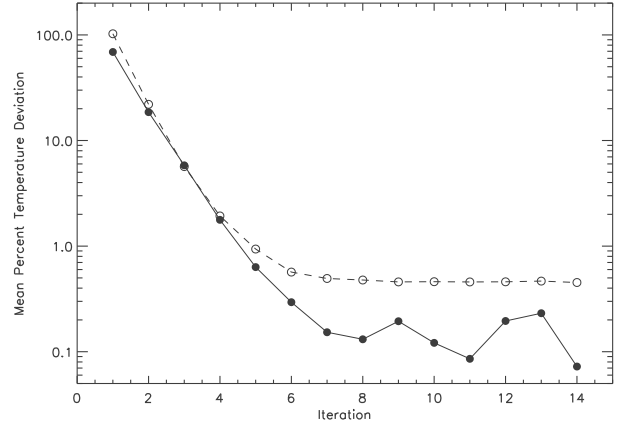}
\caption{Rate of convergence of the 1-D (filled circles, solid line)
and 2-D (open circles, dashed line) w7-like models discussed in
\S\ref{rtm:Converge}.  The figure shows, for each iteration, the mean
percent deviation of the temperature structure from that of the final
iteration 15.  Both models converge to better than 1\% in 5
iterations, with the flattening out of the curves thereafter
reflecting the level of Monte Carlo random sampling errors.  The 2-D
model has a larger number of cells, and thus the average sampling
error in each cell is larger.
\label{Fig:converge_rate}}
\end{center}
\end{figure}

\begin{figure*}[width=8.5cm,clip=true][ht]
\begin{center}
\includegraphics[height=3.4cm,clip=true]{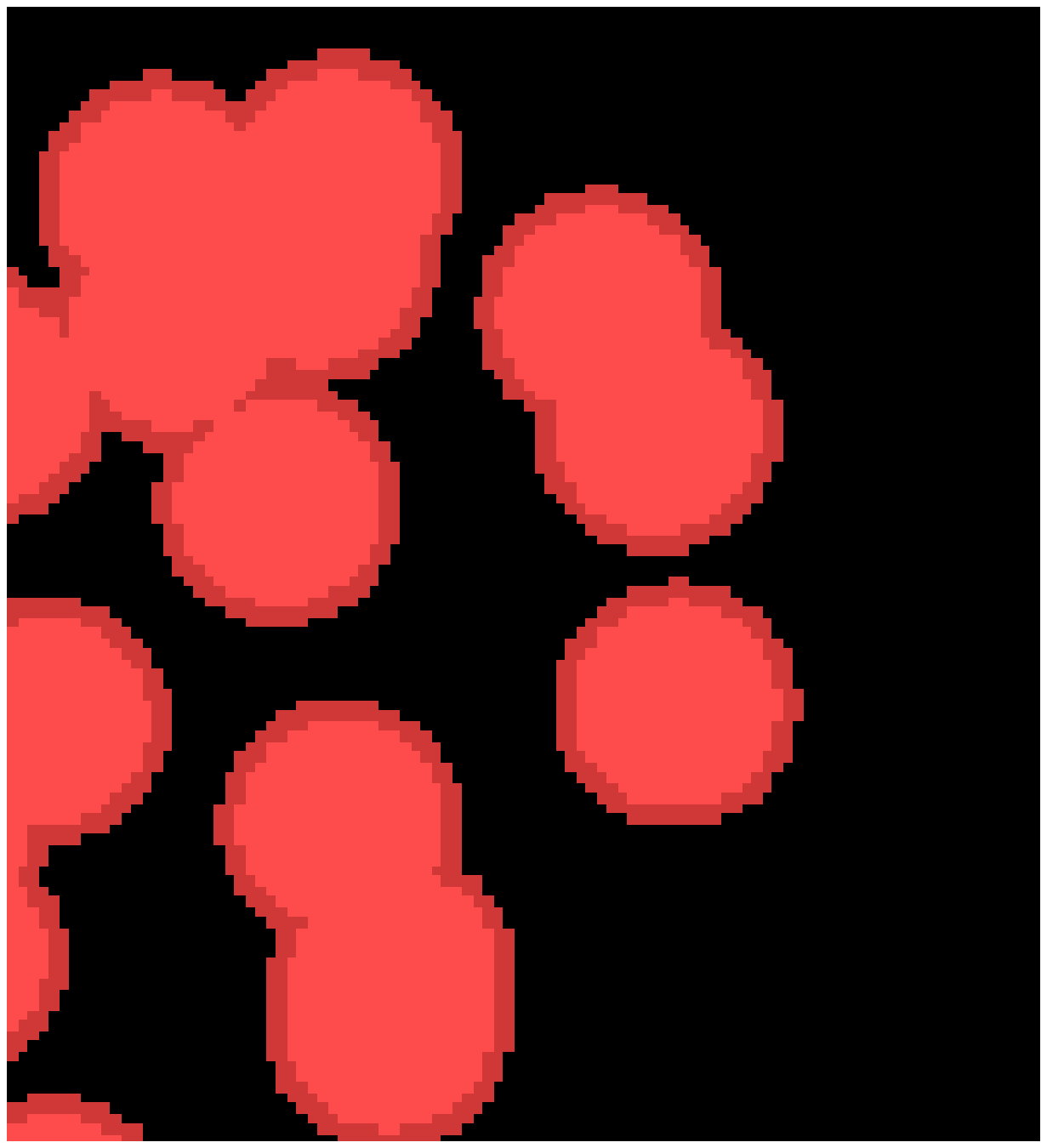}%
\includegraphics[height=3.4cm,clip=true]{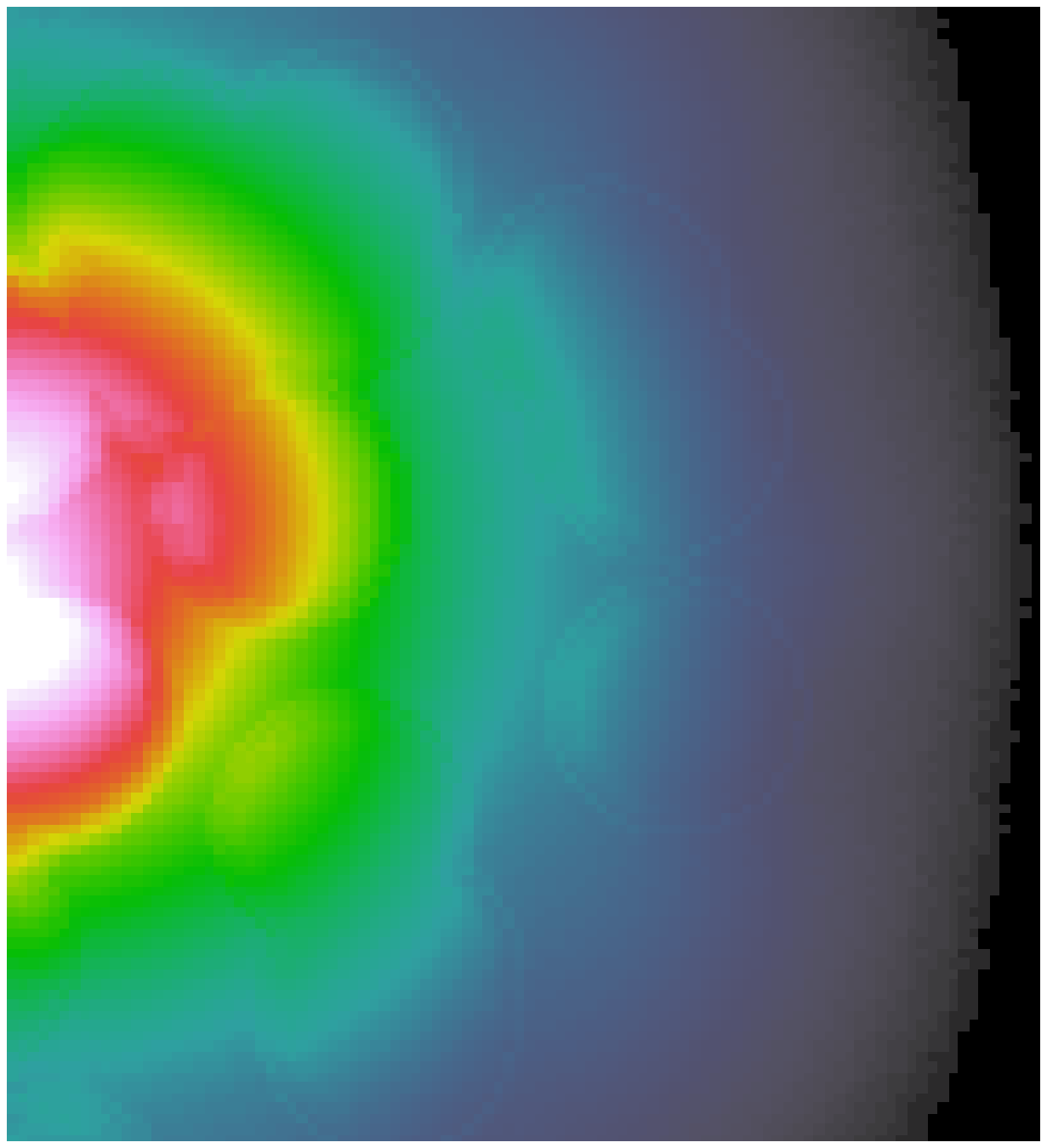}%
\includegraphics[height=3.4cm,clip=true]{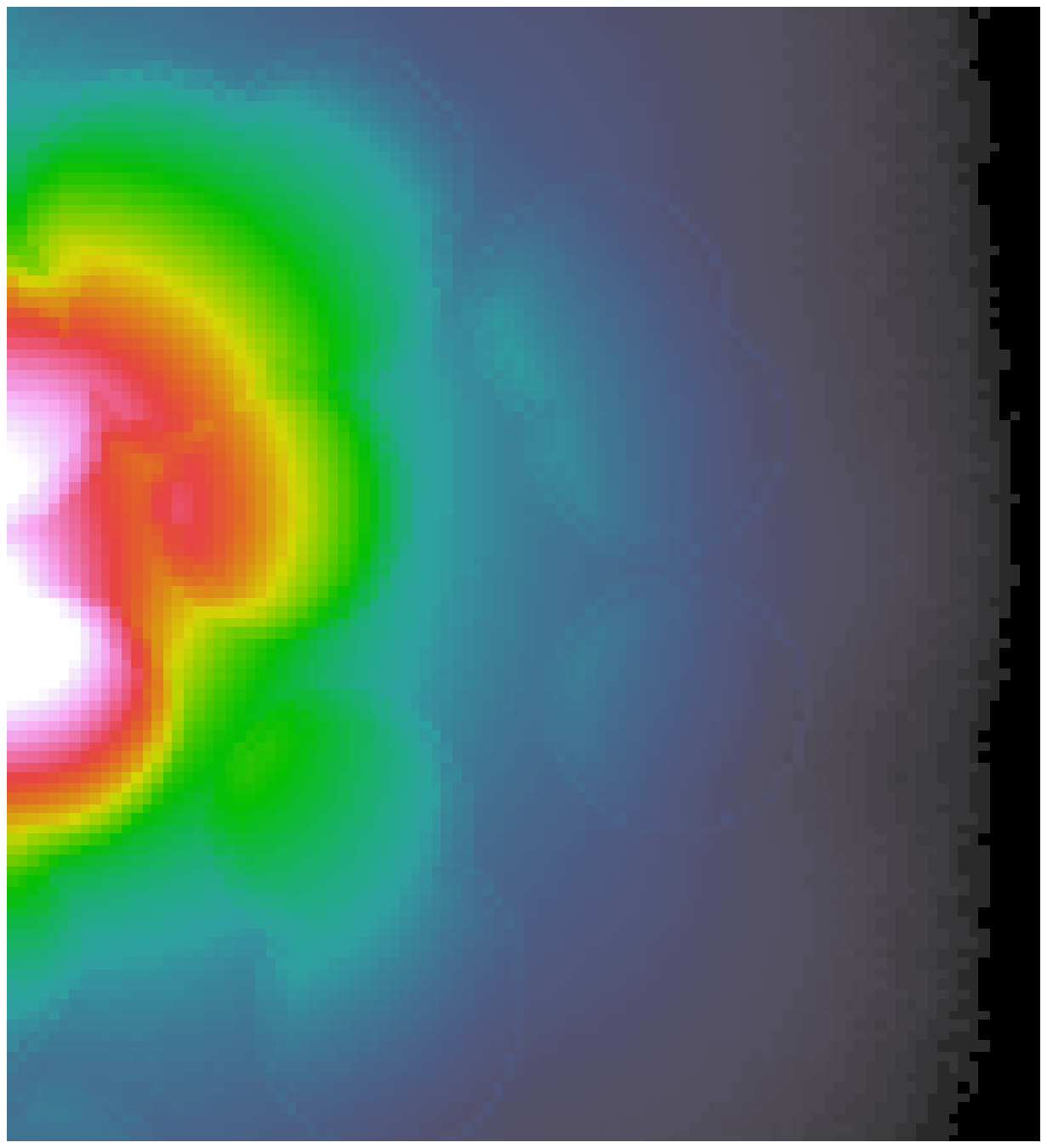}%
\end{center}
\caption{Convergence of the 2-D artificial ``clumpy'' SN~Ia model
discussed in \S\ref{rtm:Converge}.  Only the right half of the SN is
shown.  From left to right: Distribution of the \Nifs\ clumps;
Temperature structure at iteration 2; Final converged temperature
structure at iteration 15.
\label{Fig:Clump_Ia}}
\end{figure*}

An issue of particular importance in radiative transfer calculations
is the speed and stability of model convergence.  For complicated
situations -- in particular time-dependent, multi-dimensional
problems -- convergence can be the limiting factor in the practical
utility of the transfer code.

One very appealing feature of the MC approach, then, is its favorable
convergence properties. Figure~\ref{Fig:w7_converge} shows the
variation with iteration of the radial temperature structure at \texp
= 18~days for a static spectrum calculation of the 1-D w7 model.
Model parameters are those given in Table~\ref{Tab:grid}, row~4.  All
lines are treated in the expansion opacity formalism with $\epsilon =
1$.  Beginning with an isothermal atmosphere at T = 15000~K (a very
poor initial guess, in fact) convergence is stable and rapid, with the
spectrum and temperature changing negligibly after just five
iterations.  Figure~\ref{Fig:converge_rate} quantifies the convergence
rate by showing, for each iteration, the mean percent deviation of the
temperature structure from the final structure at iteration 15.  After
5 iterations, the accuracy of the temperature structure is better than
1\%.  The flattening out of the convergence curve at $0.1 \% \sim
1/\sqrt{N_p}$ reflects the level of random Monte Carlo sampling
errors, and can only be improved by increasing the number of
packets. Note that, given a more reasonable initial temperature guess
(e.g., a grey atmosphere structure) the model converges (at the 1\%
level) to the same result after only three iterations.

The rapid convergence seen in Figure~\ref{Fig:w7_converge} highlights
the utility of the constrained lambda-iteration approach developed in
\cite{Lucy_Radeq}.  In the limit that enough packets are used, the MC
transfer routine always obtains the correct radiation field at all
points, regardless of the dominance of scattering or NLTE line
processes.  Energy is strictly conserved in every packet interaction,
in contrast to difference equation solutions of the radiative transfer
in which energy is conserved only asymptotically as the iteration
procedure converges.  In the MC approach, multiple iterations are then
needed only to assure that the opacities/emissivities are consistent
with the temperature implied by energy balance.  For problems with
temperature independent opacity/emissivity, only one iteration is
required.  More generally, several iterations are needed, as each
adjustment to the opacities feeds back into an altered radiation field
structure.

Naturally, the speed of convergence hinges upon on the temperature
sensitivity of the opacity.  In the SNe~Ia example, the dominate
opacity is bound-bound line blanketing.  The contribution of an
individual line to the expansion opacity saturates for $\tau \gg 1$
(Equation~\ref{Eq:Exp_Opacity}), therefore the opacity is insensitive
to the exact optical depth of lines, depending rather on the total
\emph{number} of optically thick lines.  The variation with
temperature is therefore smooth and gradual, contributing to the rapid
and stable convergence.

These favorable convergence properties carry over to problems with
complicated multi-dimensional geometries.  We demonstrate this in
Figure~\ref{Fig:Clump_Ia} using an artificial ``clumpy'' SN~Ia model,
constructed by hand to resemble multi-D deflagration models.  The
density structure in this 2-D example was taken from the spherical w7
model, but the 0.6~\msun\ of \Nifs\ was randomly distributed in
``clumps'' (actually toruses) of velocity radius 2000~\kms.  The
``clumps'' are surrounded by a 400~\kms\ shell of silicon rich
material, and are embedded in a substrate of carbon/oxygen.  Model
parameters are given in Table~\ref{Tab:grid}, row~5.
Figure~\ref{Fig:converge_rate} shows that, despite the complicated and
irregular geometry, the model converges as quickly as the 1-D case to
better than 1\% in only 5 iterations.  Because of the larger number of
cells in the 2-D model, however, a larger number of packets are needed
to achieve adequate statistics in each cell.
Figure~\ref{Fig:Clump_Ia} shows that the final converged temperature
structure is itself highly irregular, bearing the marks of the
enhanced radioactive energy generation and ionization in the \Nifs\
clumps.  Similarly rapid convergence behavior is found for models
exhibiting global asphericity in the density contours
\citep{Kasen_hole}.

\subsection{Time Dependent Versus Stationary Spectrum Calculations}

Many SN spectral synthesis codes to date have not explicitly
included time-dependence, adopting rather a stationarity
approximation.  From a formal point of view, the assumption is
inappropriate at early times, when the photon diffusion time ($t_d$)
is long compared to the expansion time ($\texp$) of the supernova.
\Code\ allows us to test the validity of the approximation in
practice.

\Code\ calculations can be run in a time-independent, ``snapshot''
mode, such that photons are emitted according to the instantaneous
radioactive deposition function, and do not diffuse in time.  As an
example, we compute snapshot spectra of w7 at 10 days and 18 days
after explosion, for which the electron scattering diffusion times are
$t_d = 1.2$~\texp\ and 0.3~\texp, respectively.  The emergent
bolometric luminosity is a free parameter in the snapshot
calculations, which in this case was taken from the time-dependent w7
calculation discussed in \S\ref{rtm:Example}.

\begin{figure}
\begin{center}
\includegraphics[width=8.5cm,clip=true]{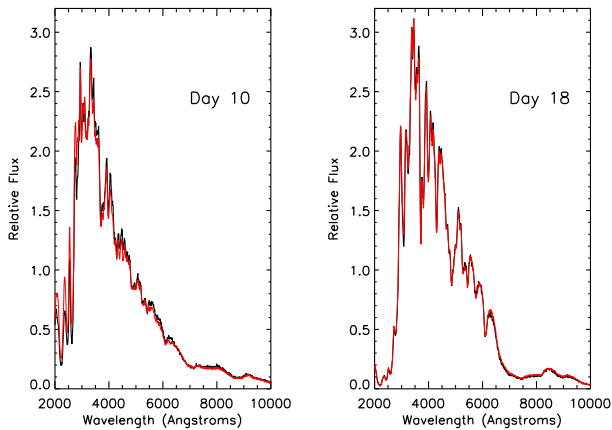}
\caption{Comparison of time-dependent spectral calculations (thick,
black lines) to time-independent ``snapshot'' calculations (thin
red,lines) for the w7 model at days \texp=10 and 18.
\label{Fig:Timedep}}
\end{center}
\end{figure}

Figure~\ref{Fig:Timedep} compares the snapshot spectra to those of the
full time-dependent calculation.  The agreement is surprisingly good
for both day~18 and day~10.  Thus, although the time-dependent
calculations include diffuse photon packets emitted at earlier epochs,
the initial spectral energy distribution of such packets has
apparently been erased during the transfer.  Wavelength redistribution
processes operate on each packet on a short enough time scale such
that the original time of emission of a photon packet is not an
important factor.

We conclude that, at least in the context 1-D SN~Ia calculations,
stationarity is a very reasonable approximation near and even prior to
maximum light.  The major limitation of the approximation, of course,
is its inability to predict the emergent bolometric luminosity at any
given time, which must therefore be included as a free parameter in
the simulation.  Stationarity may also be more questionable in
multi-dimensional polarization calculations, for which the directional
diffusion time and isotropy of the radiation field is of importance.

\subsection{Fluorescence Versus Two-Level Atom Redistribution}
\label{sec:TLA}

For SN spectral calculations, the wavelength redistribution of photons
in bound-bound transitions constitutes an essential component of the
radiative transfer.  In SNe~Ia, for example, photon packets are
typically initiated in the hot, inner regions of ejecta, where the
thermal emissivity peaks in the ultraviolet (UV).  The opacity in the
UV is large due to high density of Fe-peak lines, and the diffusion
time of UV photons exceedingly long.  Photon escape, however, is
greatly enhanced by interactions with lines, which degrade photons to
redder wavelengths, where the opacity is lower.  The primary method by
which this occurs is fluorescence (a.k.a.  line-splitting)-- i.e., a
UV photon excites an high-energy atomic transition, followed by
de-excitation via a redder transition \citep{Pinto-Eastman_II}.

Obviously a proper treatment of the wavelength redistribution in lines
is critical for calculating the lightcurves of SNe.  As previously
discussed (\S\ref{rtm:Propogate}), \Code\ allows for a direct MC
treatment of line fluorescence.  The approach, however, places
sizeable computational and memory demands on a system, and is usually
only feasible when using a restricted linelist ($\la 500,000$ lines).
A simpler, approximate method is desirable, and so in
\S\ref{rtm:Opacity} we introduced the two-level atom (TLA) expansion
opacity approach.  Here lines either scatter or ``absorb'' radiation,
depending upon the assigned redistribution probability parameter
$\epsilon$.  Absorption in a line is followed by immediate re-emission
in another line according to a thermal distribution, a process which
is designed to mimic fluorescence (the probability of true absorption
is in fact very small).

Here we explore how well a simple constant $\epsilon$ TLA approach
approximates the true wavelength redistribution calculated with
line-fluorescence treated directly.  We compute spectra and
lightcurves of the w7 model using both methods, applying a linelist of
nearly 500,000 lines from the Kurucz CD 23.  In one test calculation,
all 500,000 lines are given a direct fluorescence treatment.  In the
other test calculations, all 500,000 lines are treated using the
expansion opacity TLA formalism with different values of the
redistribution probability $\epsilon$.  We explore three values of
$\epsilon$, viz., $1.0,0.3$, and $0.01$.

\begin{figure}
\begin{center}
\includegraphics[width=8.5cm,clip=true]{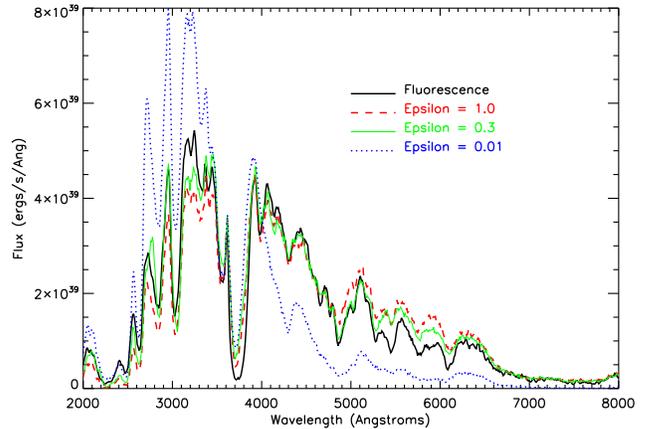}
\caption{Spectra of the w7 model (at day 18) calculated using a direct
treatment of line fluorescence (black line) compared to the TLA
approach with various redistribution probabilities (colored lines).
\label{Fig:linesplit_spectra}}
\end{center}
\end{figure}

Figure~\ref{Fig:linesplit_spectra} compares stationary w7 spectra
computed near maximum light (\texp = 18~days).  For high
redistribution probability ($\epsilon = 1$ or $0.3$) the TLA approach
in fact reasonably approximates the line-fluorescence spectrum.  In
the example, the $\epsilon = 1$ model somewhat overestimates the
redistribution, moving too much flux to the red, while the $\epsilon =
0.3$ calculation more accurately reproduces the colors.  The
scattering dominated atmosphere ($\epsilon = 0.01$), on the other
hand, dramatically fails to properly redistribute flux, leading to
unreasonable results. Similar behavior was found by
\cite{Nugent_hydro}.

\begin{figure}
\begin{center}
\includegraphics[width=8.5cm,clip=true]{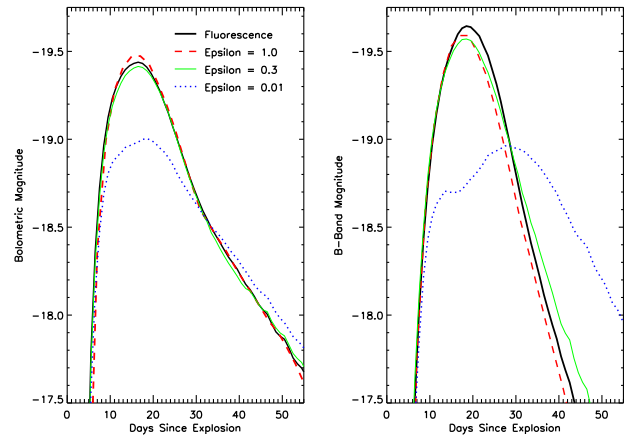}
\caption{Bolometric and B-band lightcurve of the w7 model calculated
using a direct treatment of line fluorescence (black line) compared to
the TLA approach with various redistribution probabilities (colored
lines).
\label{Fig:linesplit_LC}}
\end{center}
\end{figure}

The synthetic light curves in Figure~\ref{Fig:linesplit_LC}
demonstrate that essentially the same effect operates in the
time-dependent calculation.  For $\epsilon > 0.1$, the TLA bolometric
lightcurves are remarkably similar to the line-fluorescence
calculations.  Discrepancies are naturally larger in the monochromatic
lightcurves, which are more sensitive to the redistribution. For
example, the $\epsilon = 1$ and $\epsilon = 0.3$ B-band lightcurves
are slightly depressed at peak, as too much flux is moved to the red
(the R and I band lightcurves are correspondingly brighter). The
$\epsilon = 0.01$ atmosphere, however, gives completely unreasonable
results.  In the absence of sufficient redistribution, packets are
frozen at the high-opacity UV wavelengths. Diffusion times are long
and adiabatic losses severe.  This emphasizes the critical importance
of wavelength redistribution via fluorescence in the lightcurves of
SNe~Ia \citep{Pinto-Eastman_II}.

Overall, the TLA approach offers a useful approximation for many
lightcurve and spectral studies of interest. In the present example,
the errors in the B-band lightcurve are of order 0.1~mag, probably
comparable to other uncertainties in the transfer calculation.
Moreover, the output is not particularly sensitive to the value of
$\epsilon$, as long as $\epsilon$ is close to unity.  The
effectiveness of the TLA approach is not entirely surprising.  Given
the extreme complexity of the iron-peak element's atomic structure and
the high rate of packet-line interactions, a quasi-equilibrium is
nearly established and assuming thermal redistribution in fact fairly
well represents the actual fluorescence processes.

One could improve the TLA approximation somewhat by calibrating
$\epsilon$ for each line individually in an effort to better reproduce
the line source functions \citep[e.g.][]{Hoeflich_94D}.  We note that
in the present context, the redistribution probability for a given
atomic transition between lower level $l$ and upper level $u$ can be
approximated as
\begin{equation}
p_{\mathrm{fluor}} =  \frac{\sum_{k \ne l} \beta_{uk} A_{uk}}
{N_e \sum_k C_{uk} + \sum_k \beta_{uk} A_{uk}}. 
\label{Eq:P_Fluor}
\end{equation}
This formula expresses the probability that the radiative excitation
$l \rightarrow u$ is followed by de-excitation into a lower level
other than $l$.  For the conditions in SNe~Ia, one finds that for most
lines $p_{\mathrm{fluor}} = 0.1-1.0$.  To eliminate altogether the
free $\epsilon$ parameter occurring in our TLA approach, one could
take $\epsilon = p_{\mathrm{fluor}}$ where $p_{\rm fluor}$ is computed
using Equation~\ref{Eq:P_Fluor} for each line individually and for the
specific ejecta conditions in each spatial cell and time.  Because
$p_{\rm fluor}$ is found typically to be near unity, transfer
calculations using this approach will not in fact differ much from the
constant $\epsilon = 1$ and $\epsilon = 0.3$ calculations demonstrated
in this section.

\section{Summary and Conclusions}

In this paper we have described the computational techniques, and
demonstrated applications and verifications of the multi-dimensional
MC transfer code \Code.  We have also explored the validity of several
common approximations made in SN transfer models.  These findings will
be drawn upon in future applications of the code.

Although computational efficiency is often considered a drawback of MC
codes, the effective parallelization and favorable convergence
properties of the present techniques allow \Code\ to immediately
address complicated transfer problems using currently available
computing resources.  Moreover, as the physics included in the
radiative transfer simulation becomes increasingly complicated,
including all of multi-dimensionality, NLTE effects, polarization and
time-dependence, it is not clear that difference equation techniques
will outperform MC approaches in their execution times.

With further work, and with the advance of computing power, some of
the presently applied approximations can be relaxed.  Most outstanding
is the inclusion of an NLTE treatment of the occupation numbers,
including excitation/ionization by non-thermal electrons scattered
from radioactive gamma-rays.  Inclusion of nebular physics and cooling
by forbidden line transitions would also improve model accuracy at
late times. Eventually, coupling of the transfer code to a
multi-dimensional hydrodynamical solver would allow the calculations
to be generalized to non-homologous flows.  Fortunately, the
foreseeable improvements are readily incorporated into the MC
framework in a straightforward and intuitive manner.

\acknowledgements
We thank Leon Lucy for providing the results of his lightcurve test
calculation, David Branch for use of the SYNOW code, and Tomek Plewa
and David Jeffery for helpful comments on a draft of this manuscript.
This research used resources of the National Energy Research
Scientific Computing Center, which is supported by the Office of
Science of the U.S. Department of Energy under Contract No.
DE-AC03-76SF00098.

\appendix

\section{Gamma-Ray Transfer}

Several discussions of MC gamma-ray transfer calculations for SN can
be found in the literature, including \cite{Swartz_GRMC},
\cite{Hoeflich_GR}, \cite{Ambwani_GRMC}, \cite{Milne_GR}, and
references therein.  The important opacities for gamma-rays are
Compton scattering and photoelectric absorption (the additional
opacity due to pair-production is typically small and will be
ignored).  Because the gamma-ray energies are much greater than the
atomic binding energies, all electrons in an atom (bound and free)
contribute to the Compton opacity, which for a gamma-ray packet of
energy $E_\gamma$ is
\begin{equation}
\alphaC = \thomsonCS N K(x) \sum_i X_i Z_i,
\end{equation}
where $x = E_\gamma/m_e c^2$, $\sigma_T$ is the Thomson cross-section,
and $N$ is the total number density. The sum runs over all elements
with abundance fraction by number $X_i$ and atomic number $Z_i$.  The
dimensionless quantity $K(x)$ is the Klein-Nishina correction to the
cross-section
\begin{equation}
K(x) = 
\frac{3}{4}
\biggl[
\frac{1 + x}{x^3} 
\biggl( \frac{2x(1+x)}{1 + 2x} - \ln(1 + 2x) \biggr)
+ \frac{1}{2x} \ln (1 + 2x) - \frac{1+3x}{(1+2x)^2}
\biggr].
\end{equation}
$K(x)$ is always less than one and decreases with increasing x.

Typically Compton opacity dominates for $E_\gamma \gtrsim 50$ keV,
while photoelectric absorption dominates for lower energies.  The
photoelectric extinction coefficient is dominated by the two K-shell
electrons, and can be approximated as
\begin{equation}
\alphaP = \thomsonCS \alpha^4 8 \sqrt{2} x^{-7/2} N
\sum_i Z_i^5 X_i,
\end{equation}
where $\alpha$ is the fine-structure constant. 

Individual gamma-ray packets are emitted proportional to the local
decay rate of radioactive nuclei in one of several gamma-ray lines,
listed, for example, in \cite{Ambwani_GRMC}.  The energy of each
packet is initially $E_\gamma = E_{\rm rad,tot}/N_\gamma$, where
$E_{\rm rad,tot}$ is the total gamma-ray energy emitted from radioactive decay
and $N_\gamma$ the number of packets used in the simulation.  The
gamma-ray packets are tracked through scatterings and absorptions
through the atmosphere much as described for the optical packets in
\S\ref{rtm:Propogate}, with the propagation coming to an end when the
packet either escapes the atmosphere or is photo-absorbed in the
ejecta.

In a Compton scattering, a new direction for the gamma-ray is sampled
from the anisotropic differential cross-section
\begin{equation}
\frac{d\sigma}{d\Omega} = \frac{3\sigma_T}{16\pi} f(x,\Theta)^2
\biggl( f(x,\Theta) + f(x,\Theta)^{-1} - \sin^2\Theta \biggr),
\label{Compton_diff_cs}
\end{equation}
where $\Theta$ is the angle between incoming and outgoing gamma-ray
directions and $f(x,\Theta)$ is the ratio of incoming to outgoing
gamma-ray energy,
\begin{equation}
f(x,\Theta) = \frac{E_{\mathrm{out}}}{E_{\mathrm{in}}} = \frac{1}{1 + x (1 - \cos\Theta)}.
\label{fractional_compton_loss}
\end{equation}
The average energy lost in an interaction is given by
\begin{equation}
F(x) = 1 - \frac{1}{4\pi} \oint d\mu d \phi \frac{d\sigma}{d\Omega} f(x,\mu). 
\end{equation}
For a 1 MeV gamma-ray, $F(x) \approx 0.6$.  Thus a gamma-ray loses
almost all of its energy after just a few Compton scatterings, after
which it will be destroyed by photo-absorption.  The lost gamma-ray
energy becomes the kinetic energy of fast scattered electrons, which
are assumed to be thermalized locally through electron-electron
collisions.

The rate of energy deposition \Edepi\ in each cell $i$ and each time
step $j$ can be estimated by tallying up the gamma-ray energy lost
during each scattering or absorption event.  In this case, however,
enough packets must be used such that many interactions occur in
\emph{every} cell.  On a 3-D grid, this requires a very large number
of packets, especially at later times when interaction events become
infrequent.  Fortunately, we can derive a better estimator of \Edepi\
by considering the analytic expression for the absorbed energy,
\begin{equation}
\Edepi = \oint \int \alpha_\mathrm{abs} \Ilam d\lambda d\Omega
= 
4 \pi \int \Jlam [\alphaC(x) F(x) + \alphaP(x)] d \lambda.
\label{analytic_Edep}
\end{equation}
The mean intensity of the radiation field \Jlam\ can always be better
estimated than \Edepi\ can be by direct counting of energy loses,
because every packet passing through a cell contributes, regardless of
whether an interaction occurs.  To derive the needed estimator, we
begin with the relationship between \Jlam\ and the monochromatic
energy density $u_\lambda$ \citep{Mihalas_SA},
\begin{equation}
u_\lambda d\lambda = \frac{4\pi}{c} \Jlam d\lambda.
\end{equation}
When a packet possessing a fraction $E/E_\gamma$ of its initial
energy takes a velocity step of size \delV\ in a cell, its contribution to
$u_\lambda$ is
\begin{equation}
d u_\lambda {d \lambda} = \frac{E_\gamma}{V_i}
\biggl(\frac{E}{E_\gamma} \biggl) \biggl(\frac{\delta t}{\Delta y}
\biggr),
\end{equation}
where $V_i$ is the volume of the cell, $\Delta t$ the size of the time
slice, and $\delta t = \delV~\texp /c$. Using this equation and
Equation~\ref{analytic_Edep} gives
\begin{equation}
\Edepi = \frac{E_\gamma}{V_i~\Delta t }  \sum_k
(\delV~\texp) \frac{E}{E_\gamma} \biggl[\alphaC(x) F(x) + \alphaP(x)\biggr],
\end{equation}
where the sum over $k$ runs over every packet step that occurs inside
the cell.

Note that because the gamma-ray opacities are independent of
temperature, the gamma-ray transfer procedure need not be repeated at
every iteration.  Rather \Edepi\ need only be computed once and stored
at the beginning of a run.

\end{document}